\documentclass[twocolumn,showpacs,superscriptaddress,preprintnumbers,amsmath,amssymb]{revtex4}

\usepackage{amsmath}
\usepackage{graphicx}
\usepackage{dcolumn}
\usepackage{bm}

\begin{document}

\newcommand{\etal}{{\it et al.}}
\makeatother

\newcommand{\bra}[1]{\left\langle #1 \right|}
\newcommand{\brared}[1]{\langle #1 ||}
\newcommand{\ee}{\eta^{\ast},\eta}
\newcommand{\product}[2]{\left\langle #1 | #2 \right\rangle}

\newcommand{\kbar}{\bar{k}}
\newcommand{\ket}[1]{\left| #1 \right\rangle}
\newcommand{\ketred}[1]{|| #1 \rangle}
\newcommand{\ahat}{\hat{a}}
\newcommand{\adag}{a^{\dagger}}
\newcommand{\ahatdag}{\hat{a}^{\dagger}}
\newcommand{\Ahat}{\hat{A}}
\newcommand{\Adag}{A^{\dagger}}
\newcommand{\Ahatdag}{\hat{A}^{\dagger}}
\newcommand{\Atdag}{\hat{A}^{(\tau)\dagger}}
\newcommand{\Bdag}{\hat{B}^{\dagger}}
\newcommand{\Bhat}{\hat{B}}
\newcommand{\Bhatdag}{\hat{B}^{\dagger}}
\newcommand{\Btdag}{\hat{B}^{(\tau)\dagger}}
\newcommand{\bhat}{\hat{b}}
\newcommand{\bdag}{b^{\dagger}}
\newcommand{\cdag}{c^{\dagger}}
\newcommand{\chat}{\hat{c}}
\newcommand{\chatdag}{\hat{c}^{\dagger}}
\newcommand{\degree}{^{\circ}}
\newcommand{\sprime}{s^{\prime}}
\newcommand{\Hhat}{\hat{H}}
\newcommand{\Hhatp}{\hat{H}^{\prime}}
\newcommand{\Ihat}{\hat{I}}
\newcommand{\Jhat}{\hat{J}}
\newcommand{\hhat}{\hat{h}}
\newcommand{\fp}{f^{(+)}}
\newcommand{\fpp}{f^{(+)\prime}}
\newcommand{\fm}{f^{(-)}}
\newcommand{\Fhat}{\hat{F}}
\newcommand{\Fhatdag}{\hat{F}^\dagger}
\newcommand{\Fhatp}{\hat{F}^{(+)}}
\newcommand{\Fhatm}{\hat{F}^{(-)}}
\newcommand{\Fhatpm}{\hat{F}^{(\pm)}}
\newcommand{\Fhatdagpm}{\hat{F}^{\dagger(\pm)}}
\newcommand{\Hc}{{\cal H}}
\newcommand{\Hcp}{{\cal H}^{\prime}}
\newcommand{\Ic}{{\cal I}}
\newcommand{\It}{\widetilde{I}}
\newcommand{\ITV}{{\cal I}_{\rm TV}}
\newcommand{\Jc}{{\cal J}}
\newcommand{\jp}{j^{\prime}}
\newcommand{\Qc}{{\cal Q}}
\newcommand{\Pc}{{\cal P}}
\newcommand{\Ec}{{\cal E}}
\newcommand{\Sc}{{\cal S}}
\newcommand{\Rc}{{\cal R}}

\newcommand{\ddg}{d^{\dagger}}

\newcommand{\Nhat}{\hat{N}}
\newcommand{\Nt}{\widetilde{N}}
\newcommand{\Vt}{\widetilde{V}}
\newcommand{\nL}[1]{n_{L_{#1}}}
\newcommand{\nK}[1]{n_{K_{#1}}}
\newcommand{\nKb}{\mbox{\boldmath $n_K$}}
\newcommand{\nLb}{\mbox{\boldmath $n_L$}}

\newcommand{\mubar}{\bar{\mu}}

\newcommand{\Dc}{{\mathscr D}}
\newcommand{\Ddag}{\hat{D}^{\dagger}}
\newcommand{\dhat}{\hat{d}}
\newcommand{\Dhat}{\hat{D}}
\newcommand{\Ghat}{\hat{G}}
\newcommand{\Glambda}{G^{(\lambda)}}
\newcommand{\Gstarlambda}{G^{(\lambda)\ast}}
\newcommand{\Qhat}{\hat{Q}}
\newcommand{\Rhat}{\hat{R}}
\newcommand{\Phat}{\hat{P}}
\newcommand{\Pdag}{\hat{P}^{\dagger}}
\newcommand{\Psihat}{\hat{\Psi}}
\newcommand{\Qdag}{Q^{\dagger}}
\newcommand{\That}{\hat{\Theta}}
\newcommand{\Thatt}{\widetilde{\hat{\Theta}}}
\newcommand{\Tr}{{\rm Tr}}

\newcommand{\ktilde}{\tilde{k}}

\newcommand{\Pcirc}{\stackrel{\circ}{P}}
\newcommand{\Qcirc}{\stackrel{\circ}{Q}}
\newcommand{\Ncirc}{\stackrel{\circ}{N}}
\newcommand{\Tcirc}{\stackrel{\circ}{\Theta}}
\newcommand{\Pcircp}{\stackrel{\circ}{P^{\prime}}}
\newcommand{\Qcircp}{\stackrel{\circ}{Q^{\prime}}}
\newcommand{\Ncircp}{\stackrel{\circ}{N^{\prime}}}
\newcommand{\Tcircp}{\stackrel{\circ}{\Theta^{\prime}}}
\newcommand{\Fp}{F^{(+)}}
\newcommand{\Fm}{F^{(-)}}
\newcommand{\Rp}{R^{(+)}}
\newcommand{\Rm}{R^{(-)}}
\newcommand{\Bt}{\widetilde{B}}
\newcommand{\lambdat}{\widetilde{\lambda}}
\newcommand{\Phatt}{\widetilde{\hat{P}}}
\newcommand{\ab}{\bf a}

\newcommand{\Ab}{\mbox{\boldmath $A$}}
\newcommand{\Abdag}{\mbox{\boldmath $A$}^{\dagger}}
\newcommand{\Bb}{\mbox{\boldmath $B$}}
\newcommand{\cb}{\bf c}
\newcommand{\Db}{\mbox{\boldmath $D$}}
\newcommand{\Nb}{\mbox{\boldmath $N$}}
\newcommand{\Nbhat}{\hat{\mbox{\boldmath $N$}}}
\newcommand{\Qb}{\mbox{\boldmath $Q$}}
\newcommand{\Qhatt}{\widetilde{\hat{Q}}}
\newcommand{\Pb}{\mbox{\boldmath $P$}}
\newcommand{\phit}{\phi(t)}
\newcommand{\pdot}{\dot{p}}
\newcommand{\phix}[1]{\phi(#1)}
\newcommand{\qdot}{\dot{q}}
\newcommand{\phivib}{\phi(\eta^{\ast},\eta)}
\newcommand{\Ts}{{\cal T}}
\newcommand{\del}{\partial}
\newcommand{\eps}{\epsilon}
\newcommand{\beq}{\begin{equation}}
\newcommand{\beqa}{\begin{eqnarray}}
\newcommand{\eeq}{\end{equation}}
\newcommand{\eeqa}{\end{eqnarray}}
\newcommand{\Yb}{${}^{168}$Yb\ }
\newcommand{\Zhat}{\hat{Z}}
\newcommand{\rhodot}{\dot{\rho}}
\newcommand{\Khat}{\hat{K}}
\newcommand{\Kp}{K^{+}}
\newcommand{\Km}{K^{-}}
\newcommand{\Kz}{K^0}

\newcommand{\lb}{\bf l}
\newcommand{\sbold}{\bf s}

\newcommand{\Lp}{L^{+}}
\newcommand{\Lm}{L^{-}}
\newcommand{\Lz}{L^0}

\newcommand{\Mc}{{\cal M}}
\newcommand{\Mchat}{\hat{\cal M}}

\newcommand{\ddeta}{\frac{\partial}{\partial \eta}}
\newcommand{\ddetastar}{\frac{\partial}{\partial \eta^\ast}}
\newcommand{\etastar}{\eta^\ast}
\newcommand{\ketvib}{\ket{\phi (\etastar, \eta)}}
\newcommand{\bravib}{\bra{\phi (\etastar, \eta)}}
\newcommand{\zhateta}{\hat{z}(\eta)}
\newcommand{\zhat}{\hat{z}}
\newcommand{\oo}{\stackrel{\circ}{O}(\etastar,\eta)}
\newcommand{\oodag}{\stackrel{\circ}{O^{\dagger}}(\etastar,\eta)}
\newcommand{\oodagp}{\stackrel{\circ}{O^{\dagger\prime}}(\etastar,\eta)}
\newcommand{\oop}{\stackrel{\circ}{O^{\prime}}(\etastar,\eta)}
\newcommand{\Odag}{\hat{O}^{\dagger}}
\newcommand{\Ohat}{\hat{O}}
\newcommand{\Uinv}{U^{-1}(\etastar, \eta)}
\newcommand{\Uinvp}{U^{-1}(\etastar,\eta,\varphi,n)}
\newcommand{\U}{U(\etastar, \eta)}
\newcommand{\Up}{U(\etastar,\eta,\varphi,n)}
\newcommand{\etader}{\frac{\del}{\del \eta}}
\newcommand{\etastarder}{\frac{\del}{\del \etastar}}

\newcommand{\fb}{\mbox {\bfseries\itshape f}}
\newcommand{\SB}{\mbox {\bfseries\itshape S}}

\newcommand{\vbar}{\bar{v}}

\newcommand{\Udag}{U^{\dagger}}
\newcommand{\Vdag}{V^{\dagger}}

\newcommand{\Wc}{{\cal W}}
\newcommand{\Wcdag}{{\cal W}^{\dagger}}

\newcommand{\Xhat}{\hat{X}}
\newcommand{\Xdag}{\hat{X}^{\dagger}}

\renewcommand{\thanks}{\footnote}
\newcommand\tocite[1]{$^{\hbox{--}}$\cite{#1}}

\preprint{}

\title{Microscopic description of oblate-prolate shape mixing in proton-rich Se isotopes}

\author{Nobuo HINOHARA}
\altaffiliation[Present address: ]{Theoretical Nuclear Physics Laboratory, RIKEN Nishina Center, Wako 351-0198, Japan }
\affiliation{Yukawa Institute for Theoretical Physics, Kyoto University,
Kyoto 606-8502, Japan}
\author{Takashi NAKATSUKASA}
\affiliation{Theoretical Nuclear Physics Laboratory, RIKEN Nishina Center,
Wako 351-0198, Japan}
\author{Masayuki MATSUO}
\affiliation{Department of Physics, Faculty of Science, Niigata University,
Niigata 950-2181, Japan}
\author{Kenichi MATSUYANAGI}
\affiliation{Yukawa Institute for Theoretical Physics, Kyoto University,
Kyoto 606-8502, Japan}
\affiliation{Theoretical Nuclear Physics Laboratory, RIKEN Nishina Center,
Wako 351-0198, Japan}

\date{\today}

\begin{abstract}
The oblate-prolate shape coexisting/mixing phenomena in proton-rich 
$^{68,70,72}$Se are investigated by means of the adiabatic self-consistent 
collective coordinate (ASCC) method.
The one-dimensional collective path and the 
collective Hamiltonian describing the large-amplitude shape vibration 
are derived in a fully microscopic way. 
The excitation spectra, $B$(E2) and spectroscopic quadrupole moments are 
calculated by requantizing the collective Hamiltonian and solving the 
collective Schr\"odinger equation.
The basic properties of the coexisting two rotational 
bands in low-lying states of these nuclei are well reproduced.
It is found that the oblate-prolate shape mixing becomes weak 
as the rotational angular momentum increases. 
We point out that the rotational energy plays a crucial role 
in causing the localization of the collective wave function 
in the $(\beta, \gamma)$ deformation space.

\end{abstract}


\maketitle

\section{\label{sec:intro}Introduction}

Atomic nuclei exhibit various intrinsic shapes in their ground 
and excited states.
Coexistence of different shapes in one nucleus is 
widely observed all over the nuclear chart \cite{Wood1992101}.
Among varieties of such phenomena, much attention has been paid on 
proton-rich $N=Z$ nuclei in the $A\sim 70$ region 
where very rich shape coexistence phenomena are seen.
In this region, dramatic competition of different shapes occurs 
due to shell-structure effects; 
the oblately deformed shell gaps at $N$ or $Z=34$ and 36,
the prolately deformed shell gaps at 34 and 38, and 
the spherical shell gap at 40 \cite{Nazarewicz1985397}.

The $N=Z$ nucleus $^{68}$Se is a particularly interesting nucleus,  
because it has deformed shell gaps both at the oblate and prolate shapes. 
For this nucleus, mean-field calculations predict the oblate-prolate shape 
coexistence \cite{Takami1998242,Yamagami2001579,PhysRevC.58.R5}.
In experiment, two rotational bands were observed in low-energy excitations, 
and the ground and excited bands were interpreted to have the oblate and 
prolate deformations, respectively 
\cite{PhysRevLett.84.4064,PhysRevC.67.064318}.
For $^{70}$Se and $^{72}$Se, $B$(E2) values for the $2_1^+, 4_1^+, 6_1^+$ 
states in the ground bands were obtained by a recent life-time measurement 
\cite{ljungvall:102502}. 
These data indicate gradual change of their characters from oblate to 
prolate with increasing angular momentum; it occurs in lower angular momentum 
in $^{72}$Se compared to $^{70}$Se. The data also suggest considerable mixing 
of the oblate and prolate shapes in these low-lying states. 
We also note that candidates for the excited $0_2^+$ states have been 
known for a long time at 2011 keV in $^{70}$Se \cite{PhysRevC.24.1486} 
and at 937 keV in $^{72}$Se \cite{PhysRevLett.32.239}.

Since shape mixing is caused by large-amplitude collective motion 
connecting different shapes, 
its theoretical description is beyond the static mean-field 
approximation or the small-amplitude fluctuation about equilibrium shapes.
A difficulty in theoretical description of the shape coexisting/mixing 
phenomena is that various kinds of microscopic configurations 
associated with different shapes participate in them and thus 
quite a large number of particle-hole degrees of freedom 
are involved in the large-amplitude collective dynamics. 
Therefore, microscopic description of shape mixing is a challenging subject 
in nuclear structure theory.

Theoretical investigations on the shape coexisting/mixing phenomena 
may be divided into two categories: i.e., time-independent and time-dependent 
approaches. 
For the former, we can refer, e.g., 
the projected shell model \cite{EPJA20.133Sun},
the large-scale shell model \cite{PhysRevC.70.051301,Hasegawa200751} 
the interacting boson model \cite{al-khudair:054316} 
calculations for $^{68}$Se, 
and the excited-vampir variational calculation for $^{68,70}$Se 
\cite{Petrovici2002246,Petrovici2003396}.
For neighboring isotopes $^{72-78}$Kr, a detailed study based on 
the number and angular-momentum projected generator-coordinate method 
was recently reported \cite{bender:024312}.

A well-known approach belonging to the latter 
is the adiabatic time-dependent Hartree-Fock-Bogoliubov (ATDHFB) theory
started in late 1970's for the description of 
slow collective motions like low-frequency quadrupole vibrations 
and fissions, which exhibit strong non-linearity \cite{Ring-Schuck}.
Various versions of the ATDHFB theory have been proposed, e.g.,  
by Baranger-V\'en\'eroni \cite{Baranger1978123}, Villars \cite{Villars1977269}, 
and Goeke-Reinhard \cite{Goeke1978328}.
However, the ATDHFB approaches encountered 
some difficulties, e.g., in uniquely determining the collective path
(see \cite{Dang200093} for a review), 
so that application of the theory without introducing some additional 
approximations to the real nuclear structure has not yet been attained.

Still, challenge to develop a workable microscopic method of describing 
large-amplitude collective motions based on the ATDHF theory has been pursued.
Libert et al.\cite{PhysRevC.60.054301} developed a practical approach 
assuming the quadruple operators 
as collective coordinates and using the cranking mass. 
Quite recently, this approach was used in the discussion 
on low-lying states of $^{68,70,72}$Se \cite{ljungvall:102502}.
Using the generalized valley equation and the local RPA equation,
which are based on the ATDHFB theory, 
the shape mixing in $^{68}$Se was studied by Almehed and Walet
\cite{0954-3899-31-10-024,Almehed:2005py}. 
It is not clear, however, how the number-fluctuation degrees 
of freedom are decoupled from the large-amplitude shape vibrations 
in their work.  

On the basis of the time-dependent Hartree-Fock-Bogoliubov (TDHFB) theory, 
the self-consistent collective coordinate
(SCC) method was proposed to describe the 
large-amplitude collective motions in superconducting nuclei 
\cite{PTP.64.1294,PTP.76.372}
A new scheme of solving the SCC equations using an expansion 
in terms of the collective momentum, called adiabatic SCC (ASCC) method, 
was formulated for describing shape coexistence dynamics 
in superconducting nuclei \cite{PTP.103.959,PTP.117.451}.
It was firstly applied to the solvable multi-$O(4)$ model to demonstrate 
that it provides an efficient scheme to determine the 
collective path \cite{PTP.110.65}.

In the previous series of our works,
the ASCC method was applied to the 
oblate-prolate shape coexisting/mixing phenomena in 
$^{68}$Se and $^{72}$Kr, and the one-dimensional collective path was 
successfully determined \cite{PTP.112.363,PTP.113.129}. 
It was shown that the triaxial deformation 
plays a crucial role in the shape mixing dynamics of these nuclei. 
Furthermore, we constructed a four-dimensional collective Hamiltonian 
which can describe the coupled motion of one-dimensional collective 
vibration and the three-dimensional rotational motion of a triaxial
shape.
By requantizing the collective Hamiltonian, 
excitation spectra and quadrupole transition properties 
were evaluated \cite{PTP.119.59}.

The advantage of using the ASCC method for the description
of shape coexistence dynamics is that a few collective degrees of freedom 
relevant to the collective motion of interest 
can be  extracted self-consistently from the TDHFB phase space.
Since the collective dynamics is described in terms of
single or a few collective variables, it yields 
a clear physical interpretation of the collective dynamics.
From the collective path determined by the ASCC method, 
the direction of the collective motion can be visualized by projecting 
the collective path onto the $(\beta,\gamma)$ quadrupole deformation plane.
It is also easy to evaluate the collective inertial functions (collective mass) with respect to the $(\beta,\gamma)$ deformation coordinates. 
The obtained collective mass includes both contributions from 
the time-even and time-odd components of 
the moving mean-field \cite{Ring-Schuck}.
The time-odd contribution from the moving mean-field is ignored  
in the Inglis-Belyaev cranking formula for the collective mass,
which is widely used for the description of large-amplitude collective motions.
In the previous paper \cite{PTP.115.567}, we have shown that the
quadrupole pairing interaction enhances the collective mass
through the time-odd component of the moving mean-field.

The major purpose of this paper is to give a microscopic description, 
on the basis of the ASCC method, 
of the oblate-prolate shape mixing dynamics in proton-rich Se isotopes.
We show that the deformation degree of freedom breaking axial symmetry plays  
a crucial role in the shape mixing. Taking into account the coupling of the 
large-amplitude shape vibrations connecting the oblate and prolate shapes 
and three-dimensional rotations of the triaxial shape, 
we show that the shape mixing gradually becomes weak as the rotational 
angular momentum increases. Dynamical reason of this trend will be clarified.

This paper is organized as follows.
In Sec. \ref{sec:ascc}, the basic equations of the ASCC method
are summarized.
In Sec. \ref{sec:req}, the theoretical scheme of deriving 
the quantum collective Hamiltonian and solving the collective Schr\"odinger 
equation is described.
In Sec. \ref{sec:result}, results of the calculation for proton-rich 
$^{68,70,72}$Se isotopes are presented and discussed.
Conclusions are given in Sec. \ref{sec:conclusion}.

\section{\label{sec:ascc}The ASCC method}

In this section the basic equations of the ASCC method are summarized.
Details of their derivations are given in Ref.~\cite{PTP.103.959}.

The starting point is the time-dependent variational principle 
for a TDHFB Slater determinant representing the collective state 
$\ket{\phi(t)}$
\begin{align}
\delta \bra{\phi(t)}i\frac{\del}{\del t} - \Hhat \ket{\phi(t)} = 0, 
\label{eq:TDVP}
\end{align}
where $\Hhat$ denotes the microscopic Hamiltonian. 
In the SCC method, it is assumed that the collective motion 
could be described by a few canonical sets of collective variables.
In the present application to the shape coexistence phenomena, 
we assume that the shape dynamics can be described by 
a single collective coordinate $q$ and its canonically conjugate 
momentum $p$.
Since the system is superconducting, we also introduce 
the gauge angles $\bm{\varphi}=(\varphi^{(n)},\varphi^{(p)})$ 
together with the number fluctuation variables 
$\bm{n}=(n^{(n)},n^{(p)})$ of neutrons and protons.
Thus the TDHFB state $\ket{\phi(t)}$ 
is written in terms of these collective variables as follows.
\begin{align}
\ket{\phi(t)} = 
\ket{\phi(q,p,\bm{\varphi},\bm{n})} = 
e^{-i \sum_{\tau} \varphi^{(\tau)}\Nt^{(\tau)}}
\ket{\phi(q,p,\bm{n})}.
\end{align}
where $\Nt^{(\tau)}\equiv\Nhat^{(\tau)} - N_0^{(\tau)}$ are 
the number-fluctuation operators about the expectation values $N_0^{(\tau)}$, 
$\tau$ denoting $n$ or $p$.
Using the generalized Thouless theorem, the intrinsic state 
for the pairing rotation, $\ket{\phi(q,p,\bm{n})}$, can be written 
in terms of the moving-frame HFB state $\ket{\phi(q)}$ as
\begin{align}
\ket{\phi(q,p,\bm{n})}= e^{i\Ghat(q,p,\bm{n})} \ket{\phi(q)}, 
\end{align}
where $\Ghat(q,p,\bm{n})$ is a one-body operator.
Note that this state reduces to $\ket{\phi(q)}$ for $p=0$ and $\bm{n}=\bm{0}$; 
namely, $\ket{\phi(q,p=0,\bm{n}=\bm{0})} = \ket{\phi(q)}$.
Assuming the adiabaticity of the large-amplitude collective motion and 
the number fluctuations,
the operator $\Ghat(q,p,\bm{n})$ is expanded up to first order 
with respect to $p$ and $n^{(\tau)}$,
\begin{align}
 \Ghat(q,p,\bm{n}) =& p\Qhat(q) + \sum_{\tau} n^{(\tau)} \That^{(\tau)}(q), \\
 \Qhat(q) =& \Qhat^A(q) + \Qhat^B(q) \nonumber \\
  =& \sum_{\alpha\beta} \left(
   Q^A_{\alpha\beta}(q) \adag_\alpha \adag_\beta +
 Q^{A\ast}_{\alpha\beta}(q) a_\beta a_\alpha \right. \nonumber \\
    & \left. + Q^B_{\alpha\beta}(q) \adag_\alpha a_\beta
  \right), \label{eq:Qdef} \\
  \That^{(\tau)}(q) =& \sum_{\alpha\beta} \left(
   \Theta^{(\tau)A}_{\alpha\beta}(q) \adag_\alpha \adag_\beta + \Theta^{(\tau)A\ast}_{\alpha\beta}(q) a_\beta a_\alpha
  \right), \label{eq:Tdef}
\end{align}
where the quasiparticle creation and annihilation operators, 
$\adag_\alpha$ and $a_\alpha$, are defined with respect to 
the moving-frame HFB state $\ket{\phi(q)}$ which 
satisfies the vacuum conditions $a_\alpha \ket{\phi(q)} = 0$ for them.
Therefore these quasiparticle operators are also functions of the
collective coordinate $q$.
Note that the operator $\Qhat(q)$ contains, 
in addition to the $A$-part (the first and the second terms of 
Eq.~(\ref{eq:Qdef})), the $B$-part (the third term)
in order to satisfy the gauge invariance of the ASCC equations.
They are uniquely determined by imposing the condition 
$[\Nt^{(\tau)},\Qhat(q)]=0$ \cite{PTP.117.451}.

The collective Hamiltonian is given by 
\begin{align} \label{eq:collH}
 \Hc(q,p,\bm{n},\vec{I}) =& \bra{\phi(q,p,\bm{n})}\Hhat\ket{\phi(q,p,\bm{n})} 
 + \sum_{i=1}^3 \frac{1}{2\Jc_i(q)} I_i^2 \nonumber \\
 =& V(q) + \frac{1}{2}B(q)p^2 + \sum_\tau \lambda^{(\tau)}(q)n^{(\tau)} 
\nonumber \\
 & + \sum_{i=1}^3 \frac{1}{2\Jc_i(q)} I_i^2,
\end{align}
where 
\begin{align}
 V(q) =&  \Hc(q,p,\bm{n},\vec{I})\Big\arrowvert_{p=0,\bm{n}={\textbf 0},\vec{I}=\vec{0}}, \\
 B(q) =& \frac{\del^2 \Hc}{\del p^2}
 \Big\arrowvert_{p=0,\bm{n}={\textbf 0},\vec{I}=\vec{0}}, \label{eq:defB} \\
  \lambda^{(\tau)}(q) =& \frac{\del \Hc}{\del n^{(\tau)}}\Big\arrowvert_{p=0,\bm{n}={\textbf 0},\vec{I}=\vec{0}}, \label{eq:deflambda}
\end{align}
are the collective potential, inverse of the collective inertial
function, and the chemical potentials. 
The rotational energy term is introduced 
in order to treat the large-amplitude shape vibration and 
the three-dimensional rotation of triaxially deformed mean-field 
in a unified manner.

The moving-frame HFB equations
\begin{align}
 \delta\bra{\phi(q)}\Hhat_M(q)\ket{\phi(q)} = 0, \label{eq:mfHFB}
\end{align}
and the moving-frame QRPA equations
\begin{align}
 \delta\bra{\phi(q)}[\Hhat_M(q), \Qhat(q)] - \frac{1}{i} B(q)\Phat(q)
 \ket{\phi(q)}  = 0, \label{eq:mfQRPA1}
\end{align}
\begin{align}
 \delta\bra{\phi(q)} & [\Hhat_M(q),\frac{1}{i}\Phat(q)] -
 C(q)\Qhat(q) \nonumber \\
  &- \frac{1}{2B(q)} \left[ \left[
  \Hhat_M(q), \frac{\del V}{\del q}\Qhat(q)
\right], \Qhat(q) \right] \nonumber \\
  &- \sum_\tau \frac{\del
 \lambda^{(\tau)}}{\del q} \Nt^{(\tau)} \ket{\phi(q)} = 0, \label{eq:mfQRPA2}
\end{align}
are the basic equations which determine the collective path 
in the TDHFB phase space. They are derived by
expanding the TDHFB equation of motion (\ref{eq:TDVP}) up to second
order with respect to $p$.
Here $\Hhat_M(q)$ denotes the moving-frame Hamiltonian 
\begin{align}
\Hhat_M(q) = \Hhat - \sum_{\tau} \lambda^{(\tau)}(q) \Nt^{(\tau)} - \frac{\del V}{\del q}\Qhat(q). 
\end{align}
The operator $\Phat(q)$ is defined by
\begin{align}
 \Phat(q) \ket{\phi(q)} = i \frac{\del}{\del q}\ket{\phi(q)}. 
\end{align}
The stiffness parameter $C(q)$ is given by
\begin{align}
 C(q) = \frac{\del^2 V}{\del q^2}
 + \frac{1}{2B(q)}\frac{\del B}{\del q}\frac{\del V}{\del q}. \label{eq:defC}
\end{align}
and connected to the moving-frame QRPA frequency as $\omega^2(q)=B(q)C(q)$.

The basic equations of the ASCC method are scale invariant, 
in other words, the arbitrary scale for the collective coordinate can be 
chosen \cite{PTP.103.959}. We fix the scale by the condition $B(q)=1$. 
Note also that the method is formulated in a gauge-invariant way; 
that is, the basic equations are invariant 
under the following transformations \cite{PTP.117.451}.
\begin{align}
 \Qhat(q) \rightarrow &\Qhat(q) + \alpha^{(\tau)}\Nt^{(\tau)}, \\
 \lambda^{(\tau)}(q) \rightarrow & \lambda^{(\tau)}(q) - \alpha^{(\tau)} \frac{\del V}{\del q}(q), \\
 \frac{\del\lambda^{(\tau)}}{\del q}(q) \rightarrow & \frac{\del \lambda^{(\tau)}}{\del q}(q) - \alpha^{(\tau)}C(q).
\end{align}
Therefore, it is necessary to fix the gauge when we solve the ASCC equations.
We adopt the same gauge fixing condition 
as in Ref.~\cite{PTP.117.451}, 
which is convenient to describing shape coexisting/mixing phenomena.

In the following, we summarize the procedure of solving the ASCC equations 
starting from one of the solutions of the static HFB equations, 
which corresponds to a local minimum of the collective potential.   
The lowest frequency QRPA eigenmode at the starting HFB state $\ket{\phi(q=0)}$ 
determines the operators $\Qhat(q=0)$ and $\Phat(q=0)$.
We solve the moving-frame HFB equation (\ref{eq:mfHFB}) and 
the moving-frame QRPA equations, (\ref{eq:mfQRPA1}) and (\ref{eq:mfQRPA2}), 
off the equilibrium to obtain the solution at $q$. 
At non-equilibrium HFB states, these ASCC equations are coupled with each other,
so that the self-consistency between the moving-frame HFB state 
$\ket{\phi(q)}$ and the moving-frame QRPA mode $\Qhat(q)$ is required.
Let us assume that the solution of the ASCC equations at $q-\delta q$ 
is already known. 
We find the solution at $q$ by starting from solving 
the moving-frame HFB equation with the initial guess for 
the collective coordinate operator $\Qhat(q)$
\begin{align}
 \Qhat(q)^{(0)} = (1 - \varepsilon) \Qhat_1(q-\delta q) + \varepsilon
 \Qhat_2(q-\delta q), \label{eq:initialQ}
\end{align}
where $\varepsilon$ is a small number which mixes the lowest and 
the second-lowest solutions of the moving-frame QRPA equations at $q-\delta q$.
These two solutions usually possess different $K$-quantum numbers when  
the HFB mean field is axially symmetric.
Therefore this choice for the initial guess is crucial to find 
a symmetry-breaking solution if the previous moving-frame QRPA mode 
$\Qhat_1(q-\delta q)$ possesses the axial symmetry \cite{PTP.113.129}.
In this paper, we set $\varepsilon=0.1$ in numerical calculation.

After constructing the collective path, 
we evaluate the three rotational moments of inertia $\Jc_i(q)$. 
For this purpose, we solve the following Thouless-Valatin equations 
at every point $q$ on the collective path 
using the moving-frame HFB state $\ket{\phi(q)}$
\begin{align}
\delta \bra{\phi(q)} [\Hhat_M(q), \Psihat_i(q)]
 - \frac{1}{i}\Jc^{-1}_i(q)\Ihat_i
\ket{\phi(q)} = 0, \\
\bra{\phi(q)}[\Psi_i(q),\Ihat_i]\ket{\phi(q)}=i.
\label{eq:ascc-rot}
\end{align}
In this way, we derive the collective Hamiltonian (\ref{eq:collH}) 
from the microscopic Hamiltonian $\hat{H}$, 
which simultaneously describes the large-amplitude shape vibration 
and the three-dimensional rotation. 

\section{\label{sec:req}Requantization of the Collective Hamiltonian}

Since the collective Hamiltonian (\ref{eq:collH}) derived by the ASCC method 
is a classical one, it is necessary to requantize it 
in order to obtain collective wave functions describing shape mixing and 
discuss experimental observables such as excitation spectra and 
electromagnetic transition probabilities.

The total kinetic energy of the coupled motion of the one-dimensional
large-amplitude shape vibration and the three-dimensional rotation 
is given by
\begin{align}
 T = \frac{1}{2}B^{-1}(q)\dot{q}^2 + \sum_{i=1}^3  \frac{1}{2} \Jc_i(q) 
 \omega_i^2
   = \frac{1}{2}  \sum_{m,n} G_{mn}(q) \dot{a}_m \dot{a}_n,
\end{align}
where $\omega_i$ are angular velocities,
$\bm{\dot{a}} \equiv (\dot{q}, \omega_1, \omega_2, \omega_3)$, and the metric 
$G_{mn}(q)= \delta_{mn} (B^{-1}(q), \Jc_1(q), \Jc_2(q), \Jc_3(q))$. 
The kinetic energy term is requantized by means of the Pauli prescription:
\begin{align}
 \hat{T} =& - \frac{1}{2}\sum_{mn} |G(q)|^{-\frac{1}{2}} \frac{\del}{\del a_m}
 |G(q)|^{\frac{1}{2}} (G^{-1}(q))^{mn} \frac{\del}{\del a_n} \nonumber \\
 =& - \frac{1}{2}\frac{\del}{\del q} B(q) \frac{\del}{\del q} 
- \frac{1}{4} \frac{\del |G|}{\del q} 
  \frac{B(q)}{|G(q)|} \frac{\del}{\del q} + 
 \sum_{i=1}^3 \frac{\Ihat^2_i}{2\Jc_i(q)}, \label{eq:req-kinetic}
 \end{align}
where $|G(q)|=B^{-1}(q)\Jc_1(q)\Jc_2(q)\Jc_3(q)$ is 
the determinant of $G_{mn}(q)$.
In this paper, we take into account the second term containing the 
derivative of $|G(q)|$. 
which was ignored in our previous work \cite{PTP.119.59}.
Concerning the collective mass $B^{-1}(q)$, we can set it to unity 
without loss of generality, because it merely defines the scale for measuring 
the length of the collective path \cite{PTP.103.959}.
The three components $\Ihat_i$ of the angular momentum operator are defined 
with respect to the principal axes $(1,2,3)\equiv(x',y',z')$
associated with the moving-frame HFB state $\ket{\phi(q)}$. 
Care is needed when the collective path partially runs with axially symmetric 
shape where the moment of inertia about the symmetry axis vanishes.
We discuss this problem in subsection \ref{sec:result:path} 
with the concrete examples of 
the collective path for $^{70}$Se and $^{72}$Se. 

The collective Schr\"odinger equation is thus given 
\begin{align}
 ( \hat{T} + V(q) ) \Psi_{IMk}(q,\Omega) = E_{I,k} \Psi_{IMk}(q,\Omega),
 \label{eq:Schroedinger}
\end{align}
where $\Psi_{IMk}(q,\Omega)$ represents the collective wave function 
in the laboratory frame.
It is a function of the collective coordinate $q$ and 
the three Euler angles $\Omega$, 
and specified by the total angular momentum $I$, its projection $M$ 
on the laboratory $z$-axis,
and the index $k$ distinguishing different quantum states having 
the same $I$ and $M$.

Using the rotational wave functions ${\cal D}^I_{MK}(\Omega)$,
the collective wave functions in the laboratory frame is written as
\begin{align}
 \Psi_{IMk}(q,\Omega) =& \sum_{K=0}^I \Phi_{IKk}(q) \langle\Omega|IMK\rangle, \label{eq:collwave} \\
 \langle\Omega|IMK\rangle =&
 \sqrt{\frac{2I+1}{16\pi^2(1+\delta_{K0})}}(
 {\cal D}^I_{MK}(\Omega) + (-)^I {\cal D}^I_{M-K}(\Omega)),
\end{align}
where $\Phi_{IKk}(q)$ represents the large-amplitude vibrational motion,
and the sum in Eq.~(\ref{eq:collwave}) is restricted to even $K$.
This form of the collective wave function is analogous to that of 
the Wilets-Jean $\gamma$-unstable model \cite{PhysRev.102.788},
but the role of the triaxial deformation variable  $\gamma$ is played here by 
the collective coordinate $q$. 
Needless to say, this is a particular form in the general framework of the 
Bohr and Mottelson \cite{BMvol2}.

Normalization of the vibrational part of the collective wave functions 
is given by
\begin{align}
 \int d\tau' \sum_{K=0}^I \Phi^\ast_{IKk}(q) \Phi_{IKk'}(q) = \delta_{kk'},
\end{align}
where the volume element is 
\begin{align}
 d\tau = d\tau' d\Omega = \sqrt{|G(q)|}dq d\Omega.
 \label{eq:metric}
\end{align}

The boundary conditions for the collective Schr\"odinger equation
(\ref{eq:Schroedinger}) can be specified by projecting the collective
path obtained by the ASCC method onto the $(\beta,\gamma)$ plane
and by using the well-known symmetry properties of the Bohr-Mottelson
collective Hamiltonian \cite{Kumar1967608,PTP.119.59, BMvol2}.

\section{\label{sec:result}Shape mixing in proton-rich Se isotopes}

\subsection{\label{sec:result:parameter}Model Hamiltonian and parameters}

For the microscopic Hamiltonian, we use a version of 
the pairing-plus-quadrupole (P+Q) force model 
\cite{Bes-Sorensen,Baranger1968490} 
which includes the quadrupole-pairing in addition to the monopole-pairing 
interaction. 
Two major shells ($N_{\rm sh}=3,4$) are considered as the active model space
for neutrons and protons.
The single-particle energies are calculated using the modified oscillator 
potential \cite{Nilsson-Ragnarsson}.
As in Ref.~\cite{PTP.113.129}, 
the monopole-pairing strength $G^{(\tau)}_0$ and the quadrupole-particle-hole 
interaction strength $\chi'$ for $^{68}$Se are determined 
such that the magnitudes of quadrupole deformations and pairing gaps 
at the oblate and prolate local minima approximately reproduce  
those obtained in the Skyrme-HFB calculation 
by Yamagami et al. \cite{Yamagami2001579}.
The interaction strengths for $^{70,72}$Se are then determined 
from those of $^{68}$Se, assuming 
a simple mass number dependence $G^{(\tau)}\sim A^{-1}$ and 
$\chi'\sim A^{-\frac{5}{3}}$ \cite{Baranger1968490}.
For the quadrupole-pairing interaction strengths $G_2^{(\tau)}$, 
the self-consistent values proposed by Sakamoto and Kishimoto 
\cite{Sakamoto1990321} are evaluated from the monopole pairing interaction 
(see \cite{PTP.119.59} for details).
These values of the interaction strengths are listed in Table \ref{table:int}.

Following the conventional treatment of the P+Q model, we ignore 
the Fock term, so that, in the following, we use the abbreviation 
HB (Hartree-Bogoliubov) in place of HFB.  
The effective charges $(e_n,e_p)=(0.4e,1.4e)$ 
are used in the calculation of E2 transition matrix elements.
In numerical calculation of solving the ASCC equations, 
we use $\delta q = 0.01$.

\begin{table}
\begin{center}
\caption{Strengths of the monopole-pairing, the quadrupole particle-hole, 
and the quadrupole-pairing interactions adopted in the numerical calculation.  
The same monopole-pairing strength is used for neutrons and protons, i.e.,
we set $G_0 \equiv G_0^{(n)}=G_0^{(p)}$.
For the quadrupole-pairing interactions, 
the strengths multiplied by the oscillator length biquadrate, 
$G_2^{'(\tau)} \equiv G_2^{(\tau)}b^4$, are shown. }
\label{table:int}
\begin{tabular}{ccccc} \\ \hline \hline
           & $G_0$ (MeV) & $\chi'$ (MeV) & $G_2^{'(n)}$ (MeV) &
 $G_2^{'(p)}$ (MeV) \\ \hline
 $^{68}$Se & 0.320 & 0.248 & 0.185 & 0.185 \\
 $^{70}$Se & 0.311 & 0.236 & 0.174 & 0.184 \\ 
 $^{72}$Se & 0.302 & 0.225 & 0.161 & 0.183 \\ \hline \hline
\end{tabular}
\end{center}
\end{table}

\subsection{\label{sec:result:HB}Properties of the HB states 
and the QRPA vibrations}

The static HB solution and the QRPA calculation based on it 
provide the ASCC solution at $q=0$.
Properties of the HB mean-field and of the QRPA modes are summarized 
in Table \ref{table:HB}.
In all the three isotopes, we obtain two HB solutions possessing the oblate and 
prolate shapes.
While the magnitude of the quadrupole deformation of the oblate HB state 
depends on the neutron number rather weakly,
that of the prolate HB state significantly increases 
from $^{68}$Se to $^{72}$Se.
This trend of equilibrium deformation is consistent with what we expect from   
the deformed shell gap in the Nilsson diagram.
The oblate HB solutions are always the lowest in energy,
but the energy difference between the oblate and prolate HB local minima are
only 0.3 $\sim$ 0.6 MeV.

Concerning the QRPA vibrations in $^{68}$Se, 
the $\gamma$-vibration is the lowest frequency mode,
and the $\beta$-vibration is the second-lowest mode 
both at the oblate and prolate minima.  
The situation is different in $^{70}$Se, where 
the $\beta$-vibration is the lowest mode both at the oblate and 
prolate minima. 
In the case of $^{72}$Se, the $\gamma$-vibration is the lowest mode 
at the oblate minimum, while
the $\beta$-vibration is the lowest mode at the prolate minimum.

It is also seen in Table \ref{table:HB} that the rotational moments of 
inertia at the prolate minimum significantly increases from 
$^{68}$Se to $^{72}$Se following the increase of 
the quadrupole deformation $\beta$. 
The calculated values for $^{68}$Se and $^{70}$Se seem a little too small, 
however, compared to the values experimental data suggest.   
We plan to make a more detailed analysis about this problem in future.

\begin{table*}
\caption{\label{table:HB} 
Calculated values for the quadrupole deformations $(\beta, \gamma)$, 
the monopole pairing gaps for neutrons and protons 
$(\Delta_0^{(n)}, \Delta_0^{(p)})$, 
the potential energy $V$ measured from the lowest minimum 
of the HB equilibrium states, 
the frequencies $(\omega_\gamma, \omega_\beta)$ of 
the lowest two QRPA modes at the HB equilibrium states, 
the collective mass $M$ for the lowest QRPA mode.
The QRPA modes with $K=0$ and $K=2$ are denoted 
$\beta$ and $\gamma$, respectively,
where $K$ is the symmetry axis component of the vibrational angular momentum. 
The rotational moments of inertia $\Jc$ about the axis perpendicular 
to the symmetry axis are also shown.
}
\begin{tabular}{cccccccccc} \\ \hline\hline
 & $\beta$ & $\gamma$ & $\Delta_0^{(n)}$ (MeV) & $\Delta_0^{(p)}$ (MeV) & 
 $V$ (MeV) &
 $\omega_\gamma$ (MeV) & $\omega_\beta$ (MeV) & $M$ (MeV$^{-1}$) & $\Jc$ (MeV$^{-1}$) \\ \hline
 $^{68}$Se (ob) & 0.296 & 60$^\circ$ & 1.17 & 1.26 & 0 & 1.373 & 2.131 & 50.96 & 6.38 \\
 $^{68}$Se (pro)& 0.260 &  0$^\circ$ & 1.34 & 1.40 & 0.41 & 0.886 & 1.367 & 34.29 & 4.60 \\
 $^{70}$Se (ob) & 0.313 & 60$^\circ$ & 1.21 & 1.16 & 0 & 1.617  & 1.421 & 83.07 & 7.52 \\
 $^{70}$Se (pro)& 0.325 &  0$^\circ$ & 1.34 & 1.30 & 0.55 & 1.161  & 1.120 & 47.51 & 6.89 \\
 $^{72}$Se (ob) & 0.268 & 60$^\circ$ & 1.42 & 1.16 & 0 & 1.294 & 1.482 & 52.90 & 6.18 \\
 $^{72}$Se (pro)& 0.381 &  0$^\circ$ & 1.08 & 1.23 & 0.32 & 1.411  & 1.042 & 72.28 & 10.25\\ \hline\hline
\end{tabular}
\end{table*}

\subsection{\label{sec:result:path}Collective path}

\begin{figure*}
\begin{tabular}{ccc}
\includegraphics[width=50mm]{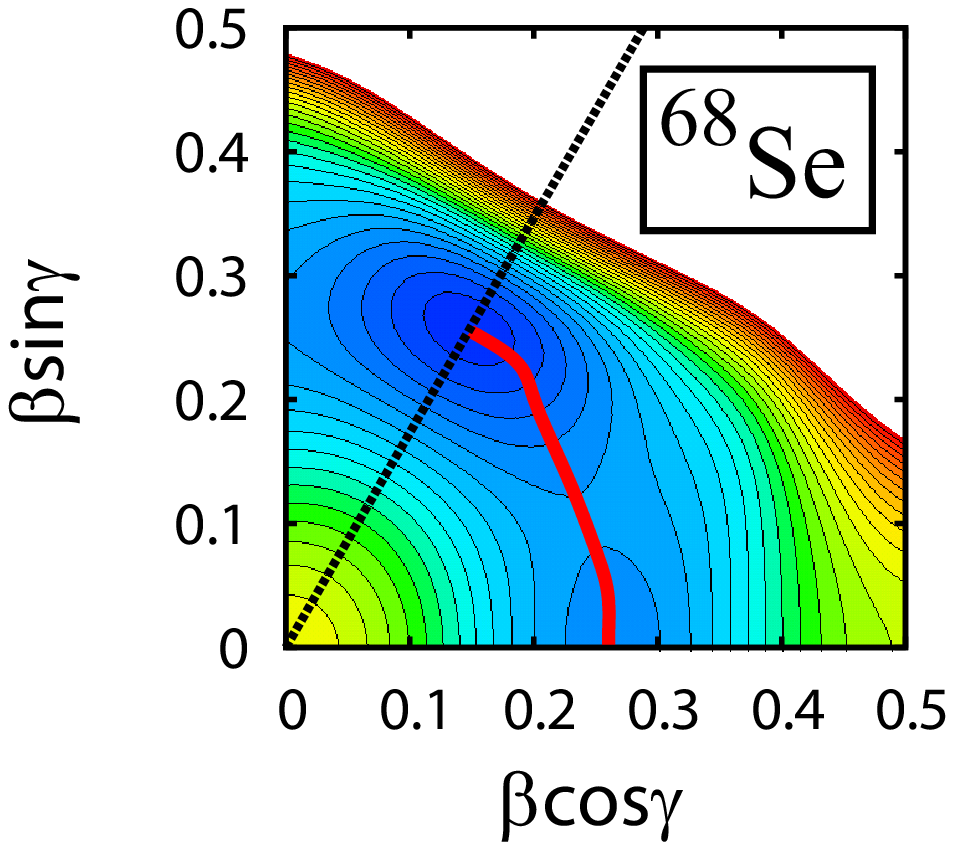} &
\includegraphics[width=50mm]{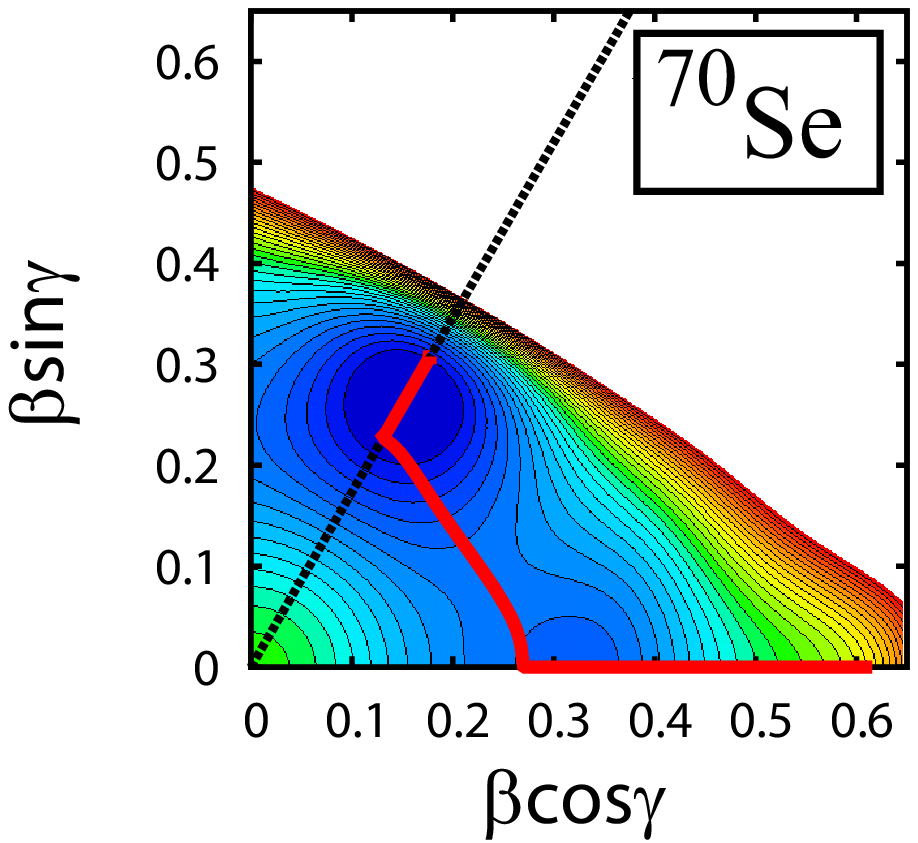} &
\includegraphics[width=50mm]{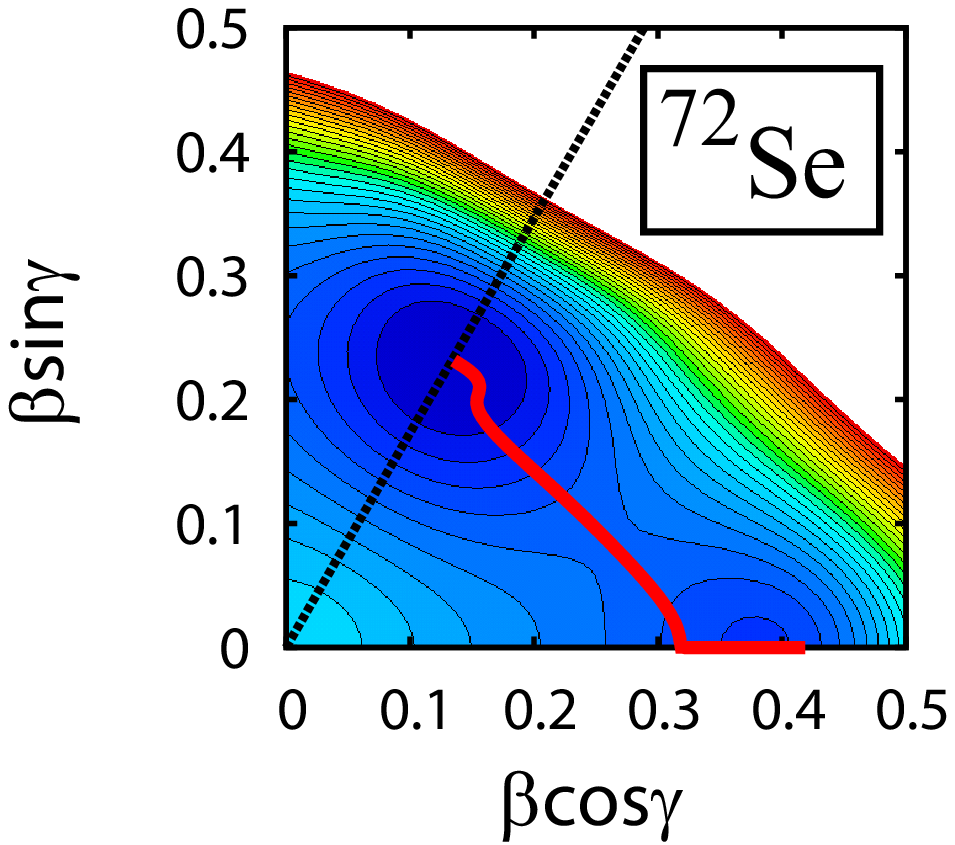}
\end{tabular}
\caption{\label{fig:path}
(Color online) The collective paths for $^{68-72}$Se obtaind by the ASCC method.
The collective path projected onto the $(\beta,\gamma)$ deformation
plane are drawn by solid lines on the potential energy surface.
The equipotential lines are drawn every 100 keV.
Note that the collective path is symmetric with respect to the
reflections about the prolate ($\gamma=0^\circ$)  and
the oblate ($\gamma=60^\circ$) axes.
The collective path going along the symmetry axis eventually
terminates at large $\beta$ when the neutron pairing gap collapses
(see the text).}
\end{figure*}

We have solved the ASCC equations and determined the collective path
choosing one of the HB solutions in Table~\ref{table:HB} in each nucleus  
and setting it as $\ket{\phi(q=0)}$. 
The results are displayed in Fig.~\ref{fig:path} where the obtained 
collective paths are drawn in the $(\beta,\gamma)$ deformation plane.

\begin{figure*}
\includegraphics[width=120mm]{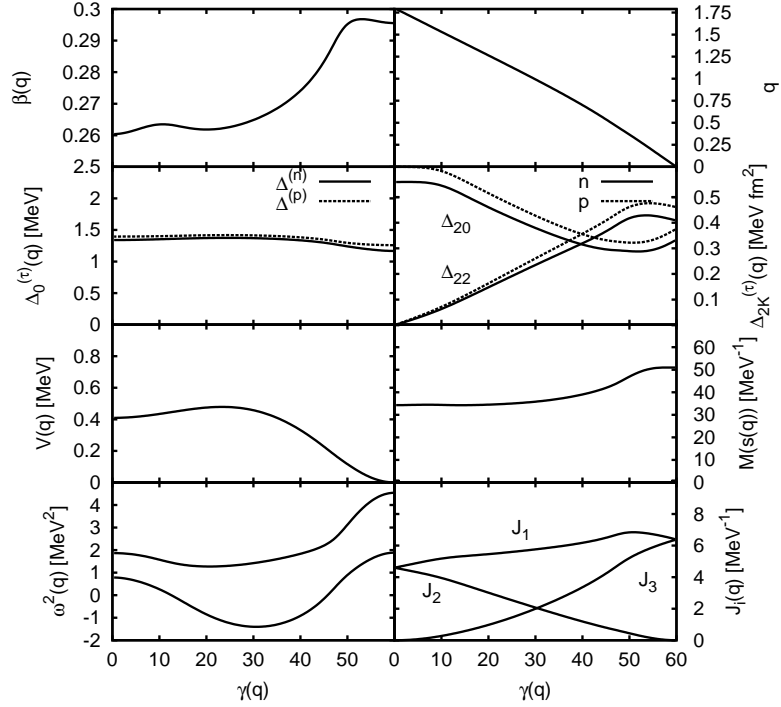}
\caption{\label{fig:68Se_ASCC}
 Results of the ASCC calculation for $^{68}$Se.
 The monopole pairing gaps $\Delta_0^{(\tau)}(q)$,
 the quadrupole pairing gaps $\Delta^{(\tau)}_{20}(q)$ and
 $\Delta^{(\tau)}_{22}(q)$,
 the collective potential $V(q)$, the collective mass $M(s(q))$,
 the rotational moments of inertia $\Jc_i(q)$, the lowest two moving-frame
 QRPA frequencies squared $\omega^2(q)$, the axial quadrupole
 deformation $\beta(q)$, and the canonical collective coordinate $q$ are
 plotted as functions of $\gamma(q)$.
}
\end{figure*}

\begin{figure*}
\includegraphics[width=120mm]{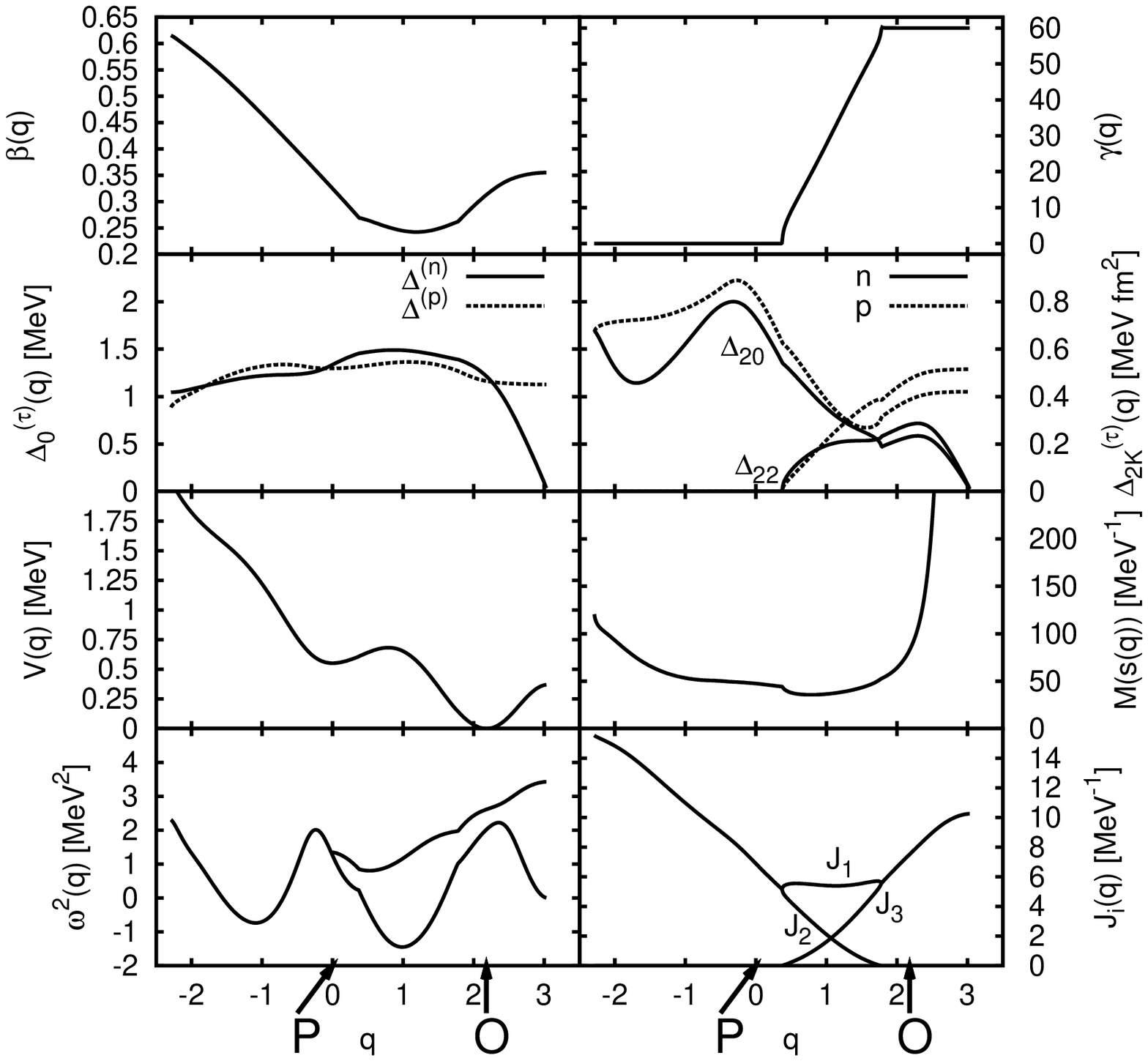}
\caption{\label{fig:70Se_ASCC}
 The same as Fig.~\ref{fig:68Se_ASCC} but for $^{70}$Se
 plotted as functions of $q$ along the collective path.
 The point $q=0$ corresponds to the prolate local minimum,
 while the oblate minimum is located at $q=2.18$.
 These positions are indicated by arrows with $P$ or $O$.
 }
\end{figure*}

\begin{figure*}
\includegraphics[width=120mm]{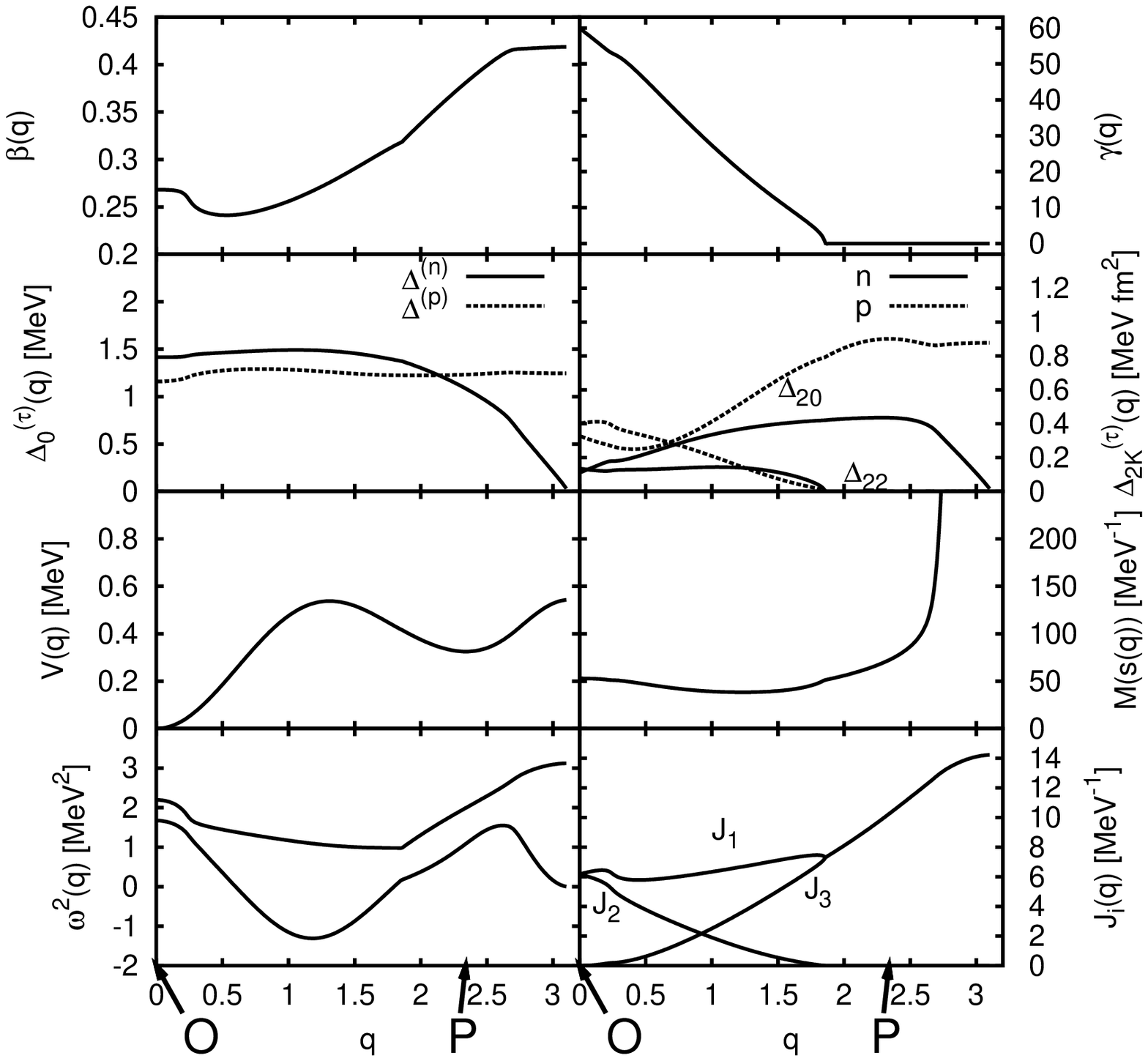}
\caption{\label{fig:72Se_ASCC}
 The same as Fig.~\ref{fig:68Se_ASCC} but for $^{72}$Se
 plotted as functions of $q$ along the collective path.
 The point $q=0$ corresponds to the oblate minimum,
 while the prolate local minimum is located at $q=2.34$.
 These positions are indicated by arrows with $O$ or $P$.
}
\end{figure*}

\subsubsection{\label{sec:result:path:68Se}$^{68}$Se}

In this nucleus, the potential barrier height is about 0.5 and 0.07 MeV 
measured from the oblate and prolate local minima, respectively. 
The oblate HB state is chosen as a starting state for solving the 
ASCC equations.
Though the collective path for $^{68}$Se is already reported 
in the previous work \cite{PTP.119.59}, 
we summarize the character of the solution of the ASCC equations
for later convenience.
The collective path starting from the oblate HB states almost follows the 
triaxial potential valley. 

In Fig.~\ref{fig:68Se_ASCC}, we see that the deformation $\beta$ almost stays 
constant during when the triaxial deformation $\gamma$ changes 
from $60^\circ$ to $0^\circ$ along the collective path.
This clearly indicates that the triaxial degree of freedom plays much more 
important role than the axial degree of freedom in $^{68}$Se.
The $\gamma$-dependence of the calculated moments of inertia exhibits 
a behaviour similar to the irrotational moments of inertia; 
two of them coincide at the axially symmetric limit 
while the largest moment of inertia is about the intermediate axis. 

In Fig.~\ref{fig:68Se_ASCC},
the collective mass defined as a function of the 
geometrical length $ds = \sqrt{ d\beta^2 + \beta^2 d \gamma^2 }$ 
in the ($\beta,\gamma$) plane,
\begin{align}
 M(s(q)) = B^{-1}(q) 
\left\{ 
\left(
\frac{d\beta}{dq}\right)^2 + \beta^2(q)\left(\frac{d\gamma}{dq}\right)^2 \right\}, 
\end{align}
is also presented. 
As mentioned in Sec. \ref{sec:req}, we can set $M(q)=B(q)^{-1} = 1~$MeV$^{-1}$ 
using the units where  $\hbar=1$ and the collective variables $(q, p)$ are non-dimensional.

\subsubsection{\label{sec:result:path-70Se}$^{70}$Se}

In this nucleus, the potential barrier height is about 0.7 and 0.1 MeV 
measured from the oblate and prolate local minima, respectively. 
The collective path is obtained starting from the prolate HB state.
The two HB local minima are connected by the one-dimensional path.
Since the QRPA mode with the lowest frequency at the prolate shape 
has $\beta$-vibrational character with $K=0$, 
the collective path starting from the prolate HB state
goes along the axial symmetry axis in the beginning.
As seen in Figs.~\ref{fig:path} and \ref{fig:70Se_ASCC}, 
at $q\simeq0.4~(\beta\simeq0.27)$, 
the collective path deviates from the $\gamma=0^\circ$ line 
due to the character change of the lowest mode from $K=0$ to $K=2$.
To describe such a dynamical breaking of the axial symmetry 
taking place along the collective path, it is crucial to use 
Eq.~(\ref{eq:initialQ}) as an initial trial for self-consistently 
determining the collective coordinate operator $\hat{Q}(q)$.
The collective path encounters a similar avoided crossing of 
the moving-frame QRPA modes 
at $q\simeq1.8$ (the oblate side with $\beta\simeq0.27$). 
Then, the triaxial path again changes its direction and go along 
the $\gamma=60^{\circ}$ line.
This kind of dynamical breaking of the axial symmetry was previously 
reported in the analysis of the collective path for 
$^{72}$Kr \cite{PTP.113.129, PTP.119.59}.
At the oblate side of the collective path,
the $\beta$-vibrational degrees of freedom strongly couples with 
the pairing-vibrational degree of freedom of neutrons, 
and it ends at a large $\beta$ point where the neutron pairing gap collapses.
When approaching this point, the collective mass diverges.

As the rotation about the symmetry axis disappear, 
the moments of inertia $\Jc_3(q)$ and $\Jc_2(q)$ should be dropped in the 
determinant $|G(q)|$ of the metric $G_{mn}(q)$ when the collective path 
runs along the $\gamma=0^{\degree}$ and $\gamma=60^{\degree}$ lines, 
respectively. To avoid discontinuity at the point where the collective path 
starts to deviate from the symmetry axis, we use 
$|G(q)|=B^{-1}(q)\Jc_1(q)$ for the whole region of the collective path  
of this nucleus.

In the region of prolate shape with $\beta > 0.34$, the lowest two QRPA modes 
with $\beta$ and $\gamma$ characters approximately degenerate in energy and 
compete (see Fig.~\ref{fig:70Se_ASCC}).  
In such a situation, it may be appropriate to introduce 
two collective coordinates. In the present calculation, however, 
we have solved the ASCC equations 
in this region assuming that the collective path continues to go 
along the $\gamma=0^{\circ}$ axis. 
We shall attempt to introduce two collective coordinates 
in the same framework of the ASCC method in future.

\subsubsection{\label{sec:result:path-72Se}$^{72}$Se}

In this nucleus, the potential barrier height is about 0.5 and 0.3 MeV 
measured from the oblate and prolate local minima, respectively. 
The collective path is determined starting from the oblate HB state. 
As seen in Figs.~\ref{fig:path} and \ref{fig:72Se_ASCC}, 
the collective path directs to the triaxial region 
because the character of the lowest QRPA mode at the oblate minimum 
is $\gamma$-vibrational.
At $q\simeq0.2$ in the triaxial region, the collective path curves 
due to an interplay of the lowest two moving-frame QRPA modes.
The collective path reaches the prolate side at $q\simeq1.6~(\beta\simeq0.32)$. 
Then the path changes its direction to the $\gamma=0^\circ$ line.
Thus, the oblate and prolate local minima are connected by a single 
collective coordinate.
After passing through the prolate minimum at $q\simeq2.1~(\beta\simeq0.38)$,
it continues to go along the $\gamma=0^\circ$ line and 
finally terminates at $\beta\simeq0.42$ where the neutron pairing gap collapses.
Correspondingly, the collective mass increases with increasing $\beta$ and 
finally diverges.  

As discussed above for $^{70}$Se, 
the moment of inertia $\Jc_3(q)$ should be dropped in the 
determinant $|G(q)|$ when the collective path 
runs along the $\gamma=0^{\degree}$ line. 
Since the collective path for $^{72}$Se does not run along   
the $\gamma=60^{\degree}$ line at all, 
we use $|G(q)|=B^{-1}(q)\Jc_1(q)\Jc_2(q)$ 
for the whole region of the collective path of this nucleus.  

\subsection{\label{sec:result:energy}
Shape mixing, excitation spectra, quadrupole transitions and moments}

We have calculated collective wave functions solving 
the collective Schr\"odinger equation  (\ref{eq:Schroedinger}) 
and evaluated excitation spectra, quadrupole transition probabilities 
and spectroscopic quadrupole moments. 
Below we discuss these results 
denoting the eigenstates belonging to 
the ground and excited bands as  $0_1^+, 2_1^+, 4_1^+, 6_1^+$ and  
$0_2^+, 2_2^+, 4_2^+, 6_2^+$, respectively. 

\subsubsection{\label{sec:result:energy:68Se}$^{68}$Se}

\begin{figure*}
\includegraphics[width=100mm]{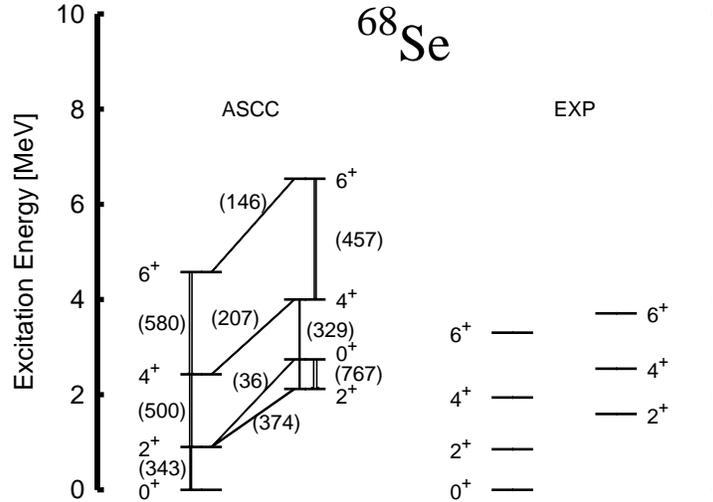}
\caption{\label{fig:68Se_energy}
Theoretical ({\it left}) and experimental ({\it right}) excitation spectra
and $B$(E2) values for low-lying states in $^{68}$Se.
Only $B$(E2)'s larger than 1 Weisskopf unit are shown in units of $e^2$ fm$^4$.
Experimental data are taken from \cite{PhysRevLett.84.4064,PhysRevC.67.064318}.
}
\end{figure*}

\begin{figure}
\includegraphics[width=90mm]{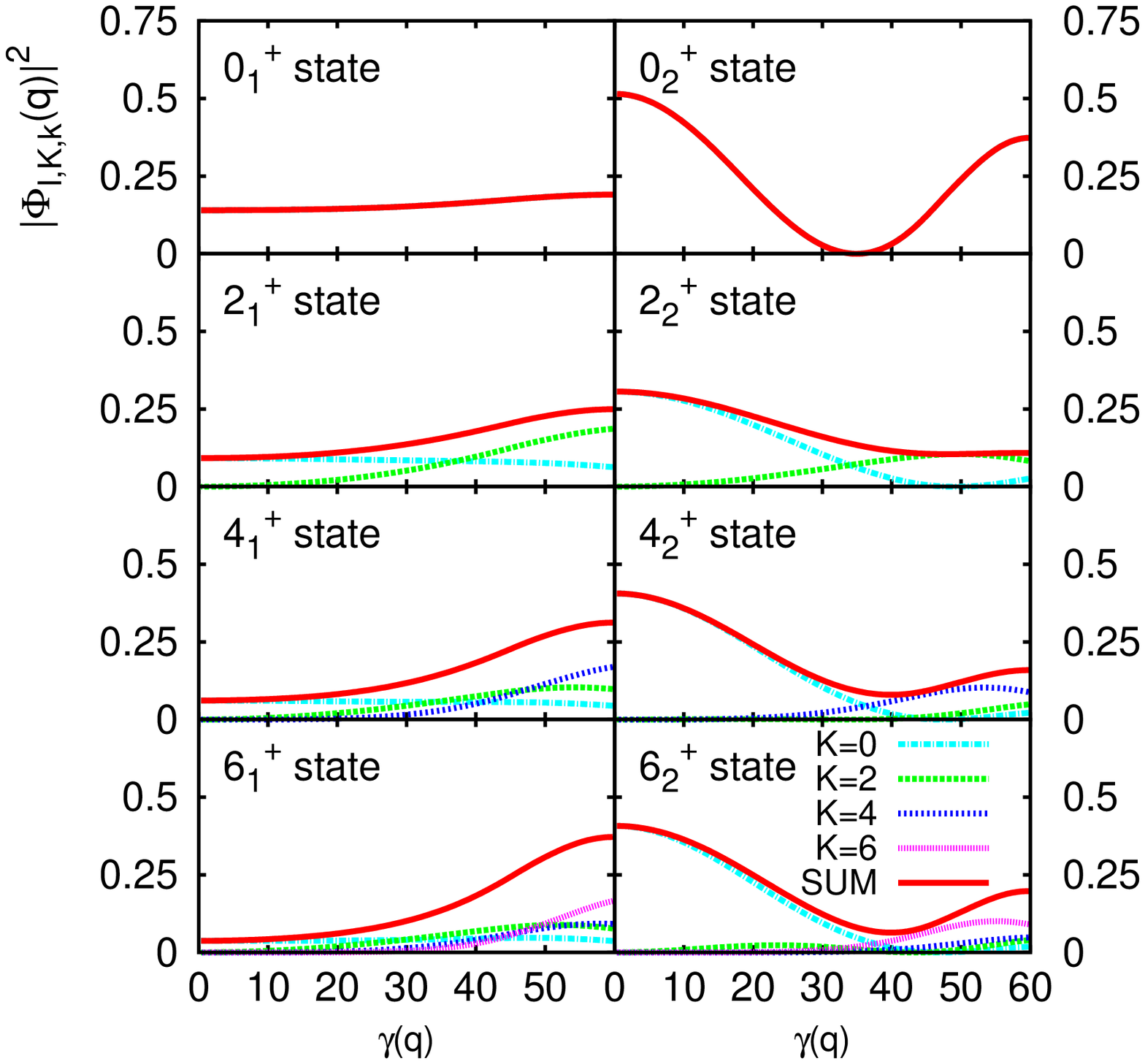}
\caption{ \label{fig:68Se_wave}
(Color online)
Vibrational wave functions $\Phi_{IKk}(q)$ squared 
of the lowest ({\it left}) and the second-lowest states ({\it right})
for each angular momentum in $^{68}$Se. 
In each panel, different $K$ components of 
the vibrational wave function and the sum of them are 
plotted as functions of $\gamma(q)$.}
\end{figure}

In Fig.~\ref{fig:68Se_energy}, excitation spectrum 
and $B$(E2) values calculated for $^{68}$Se 
are displayed together with experimental data. 
It is seen that the calculation yields two bands 
which exhibit significant deviations from the regular rotational pattern. 
We note, in particular, that the calculated $0_2^+$ state is located 
above the $2_2^+$ state.
This is consistent with the available experimental data 
where the $0_2^+$ state has not yet been found.
We see in Fig.~\ref{fig:68Se_wave} that the vibrational wave functions 
of the $0_1^+, 0_2^+, 2_1^+$ and $2_2^+$ states spread over 
the whole extent of $\gamma$ from the oblate to the prolate shapes. 
This result of calculation is reasonable considering the very low  
potential barrier along the triaxial collective path, 
as we have already seen in Fig.~\ref{fig:68Se_ASCC}. 
The unusual behavior of the excited $0^+$ state noted above  
suggests that the low-lying states in $^{68}$Se are in 
an intermediate situation between the oblate-prolate shape coexistence and 
the Wilets-Jean $\gamma$-unstable model \cite{PhysRev.102.788}. 
In fact, we can find a pattern in the calculated E2-transition probabilities, 
which is characteristic to the $\gamma$-unstable situation; for instance, 
$B$(E2; $6_2^+ \to 6_1^+$), $B$(E2; $4_2^+ \to 4_1^+$) and $B$(E2; $2_2^+ \to 2_1^+$)
are much larger than   
$B$(E2; $6_2^+ \to 4_1^+$), $B$(E2; $4_2^+ \to 2_1^+$) and $B$(E2; $2_2^+ \to 0_1^+$). 
This point will be discussed with a more general perspective 
in a future publication \cite{Sato-inprep}. 
It is quite interesting to notice that the shape mixing becomes weak 
as the angular momentum increases, and the collective wave functions of
the $4^+$ and $6^+$ states tend to localize in the region near 
either the oblate or the prolate shape.  
Namely, it becomes possible to characterize 
the $4^+$ and $6^+$ states as oblate-like or prolate-like. 

\subsubsection{\label{sec:result:energy:70Se}$^{70}$Se}

\begin{figure*}
\includegraphics[width=100mm]{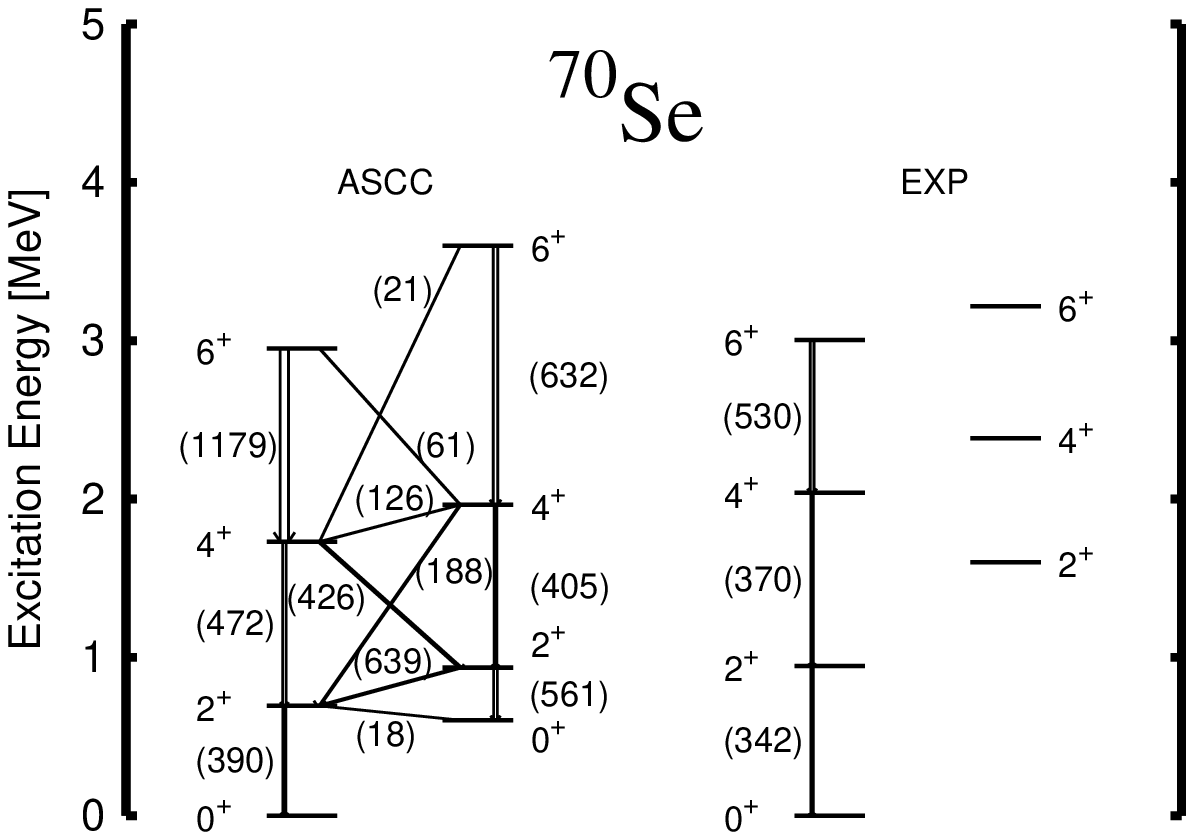}
\caption{\label{fig:70Se_energy}
The same as Fig.~\ref{fig:68Se_energy} but for $^{70}$Se.
Experimental data are taken from \cite{0954-3899-28-10-307,ljungvall:102502}. 
A candidate of the $0_2^+$ state is suggested in experiment 
\cite{PhysRevC.24.1486} at about 2 MeV, although it is not drawn.}
\end{figure*}

\begin{figure}
\includegraphics[width=90mm]{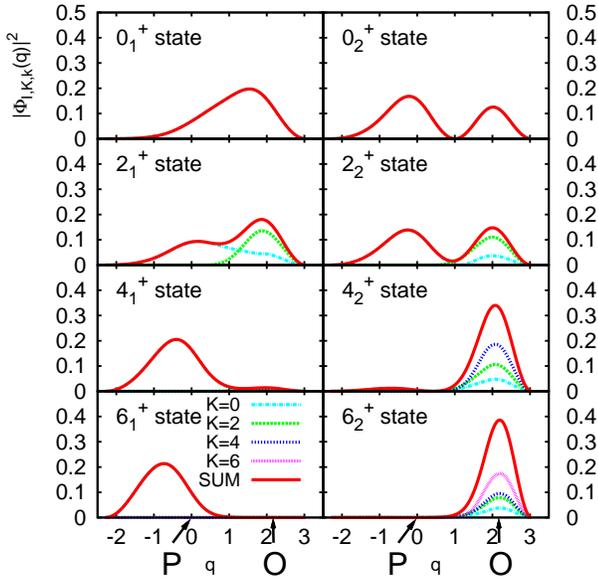}
\caption{ \label{fig:70Se_wave}
(Color online)
Same as Fig.~\ref{fig:68Se_wave} but for $^{70}$Se. 
The vibrational collective wave functions squared are
plotted as functions of $q$.
The arrows indicate the positions of the oblate ($O$) and the 
prolate ($P$) minima.
}
\end{figure}

Calculated and experimental excitation spectra and $B$(E2) values 
for $^{70}$Se are displayed in Fig.~\ref{fig:70Se_energy}.
The excitation energies of the ground band is well reproduced. 
The calculated $B$(E2;$2_1^+\rightarrow 0_1^+$) value 390 $e^2$ fm$^4$
is also in reasonable agreement with the experimental data 342 $e^2$ fm$^4$.
The calculated E2-transition probabilities exhibit a pattern somewhat 
different from that of $^{68}$Se; for instance, we see significant cross 
talks among the  $2_1^+, 2_2^+, 4_1^+$ and $4_2^+$ states. 
The vibrational wave functions of the $0_1^+, 0_2^+, 2_1^+$ and $2_2^+$ states 
displayed in Fig.~\ref{fig:70Se_wave} show strong oblate-prolate shape mixings.
In contrast, the $4_1^+$ and $6_1^+$ ($4_2^+$ and $6_2^+$) states are rather 
well localized around the prolate (oblate) shape. 
Thus, the characteristic cross talks of the E2 transition strengths 
mentioned above are associated with the significant change in 
localization properties of the vibrational wave functions 
between angular momenta 2 and 4.  
Experimental data for such inter- and intra-band $B$(E2) values will certainly 
serve as a very good indicator of the shape mixing. 

\subsubsection{\label{sec:result:energy:72Se}$^{72}$Se}

\begin{figure*}
\includegraphics[width=100mm]{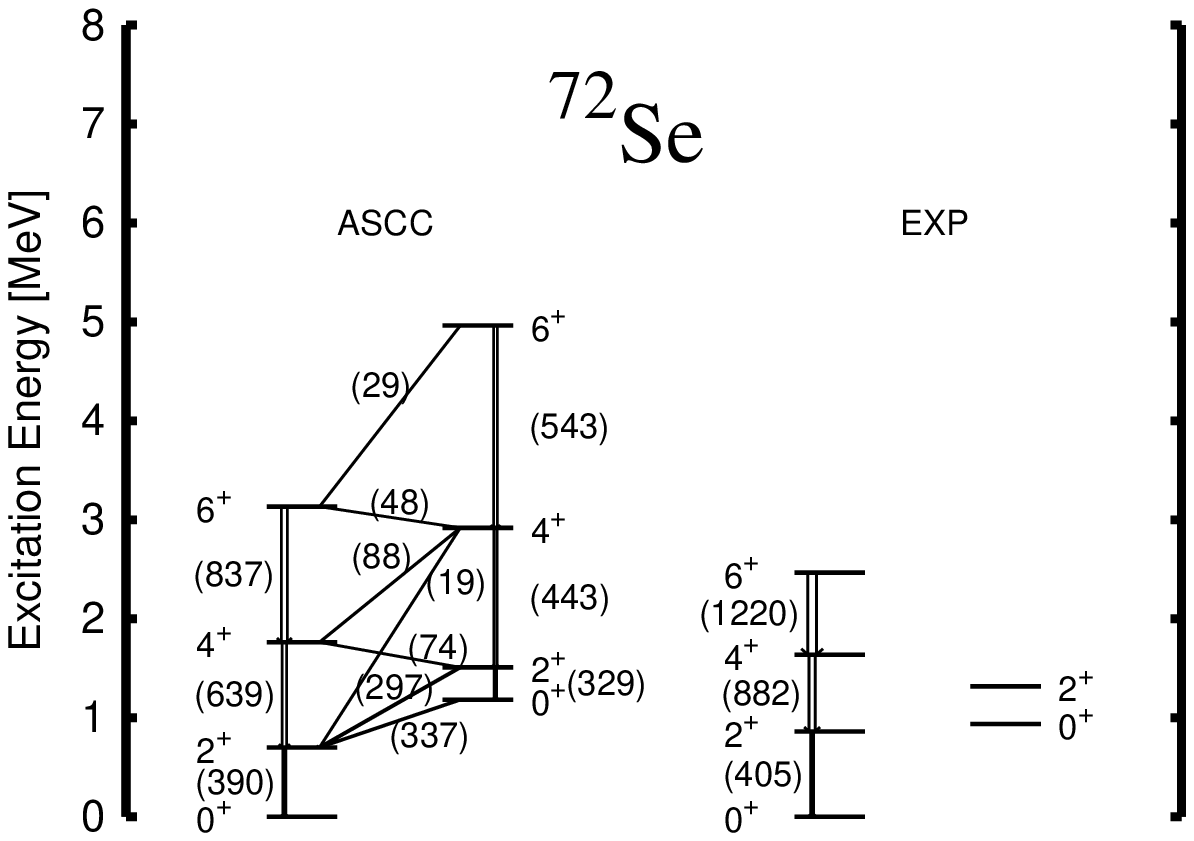}
\caption{ \label{fig:72Se_energy}
The same as Fig.~\ref{fig:68Se_energy} but for $^{72}$Se.
Experimental data are taken from \cite{PhysRevC.63.024313,ljungvall:102502}.}
\end{figure*}

Calculated excitation spectrum and $B$(E2) values for $^{72}$Se are 
shown in Fig.~\ref{fig:72Se_energy} together with experimental data.
It is seen that the experimental spectrum is reproduced fairly well. 
The calculated $B$(E2;$2_1^+\rightarrow 0_1^+$) value 390 $e^2$ fm$^4$
is also in good agreement with the experimental data 405 $e^2$ fm$^4$.
We see that the calculated inter-band E2 transitions,
$B$(E2; $4_2^+ \to 4_1^+$), $B$(E2; $4_2^+ \to 2_1^+$), $B$(E2; $2_2^+ \to 2_1^+$) 
and $B$(E2; $4_1^+ \to 2_2^+$), are reduced from those in $^{70}$Se, 
except for $B$(E2; $0_2^+ \to 2_1^+$).

\begin{figure}
\includegraphics[width=90mm]{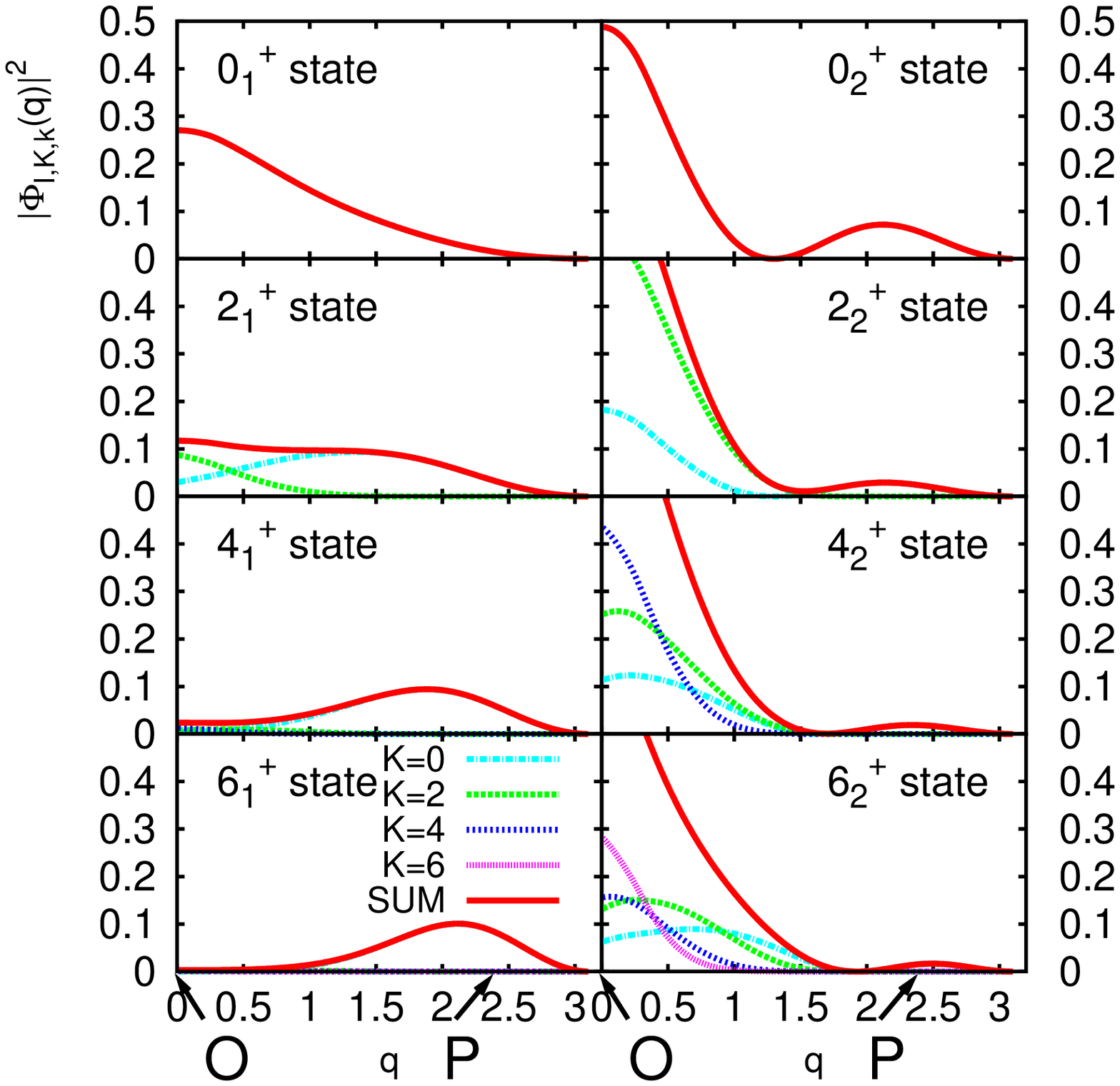}
\caption{ \label{fig:72Se_wave}
(Color online)
Same as Fig.~\ref{fig:68Se_wave} but for $^{72}$Se.
The vibrational wave functions squared are plotted as functions of $q$.
The arrows indicate the positions of the oblate ($O$) and the 
prolate ($P$) minima.
}
\end{figure}

The vibrational wave functions are displayed in Fig.~\ref{fig:72Se_wave}.
Similarly to $^{70}$Se, the $0_1^+$ wave function widely spreads over 
the triaxial region. 
It takes the maximum at the oblate shape, but extends to the prolate region.  
The $2_1^+$ wave function also extends the whole region of $\gamma$. 
In the ground band, the prolate character develops  with increasing 
angular momentum, as clearly seen in the wave functions of the 
$4_1^+$ and $6_1^+$ states. 

Dynamical reason why the prolate character of the ground band develop with 
increasing angular momentum may be understood in terms of the competition 
between the potential and kinetic energies as function of the 
collective coordinate $q$. We find that the rotational energy term plays 
a particularly important role. Because the quadrupole deformation 
$\beta \simeq 0.38$ at the prolate local minimum is much larger than that 
($\beta \simeq 0.27$) at the oblate minimum, the moment of inertia 
$\Jc \simeq 10.3$ MeV$^{-1}$ at the former is appreciably larger 
than  $\Jc \simeq 6.2$ MeV$^{-1}$ at the latter. 
The difference between the rotational energies at the prolate and oblate minima 
is easily evaluated to be about 0.19, 0.64, 1.35 MeV for the 
$2^+, 4^+, 6^+$ states, respectively.
Therefore, the prolate shape is favoured in order to reduce the 
rotational energy. As the rotational angular momentum increases, 
this rotational effect becomes more important and overcomes 
the small disadvantage in the potential energy. 
In contrast, for the $0^+$ ground state where the rotational effect is absent, 
the vibrational wave function takes the maximum at the oblate minimum. 
For the $2_1^+$ state, the difference of the rotational energies, 
about 0.19 MeV, is slightly smaller than that of the potential energies,
about 0.32 MeV, so that its collective wave function exhibits 
a transitional character from oblate-like to prolate-like. 
For the excited states, the vibrational wave functions 
possess the dominant bumps  
around the oblate shape, exhibiting at the same time the second bumps 
around the prolate shape. 

\begin{figure}
\includegraphics[width=70mm]{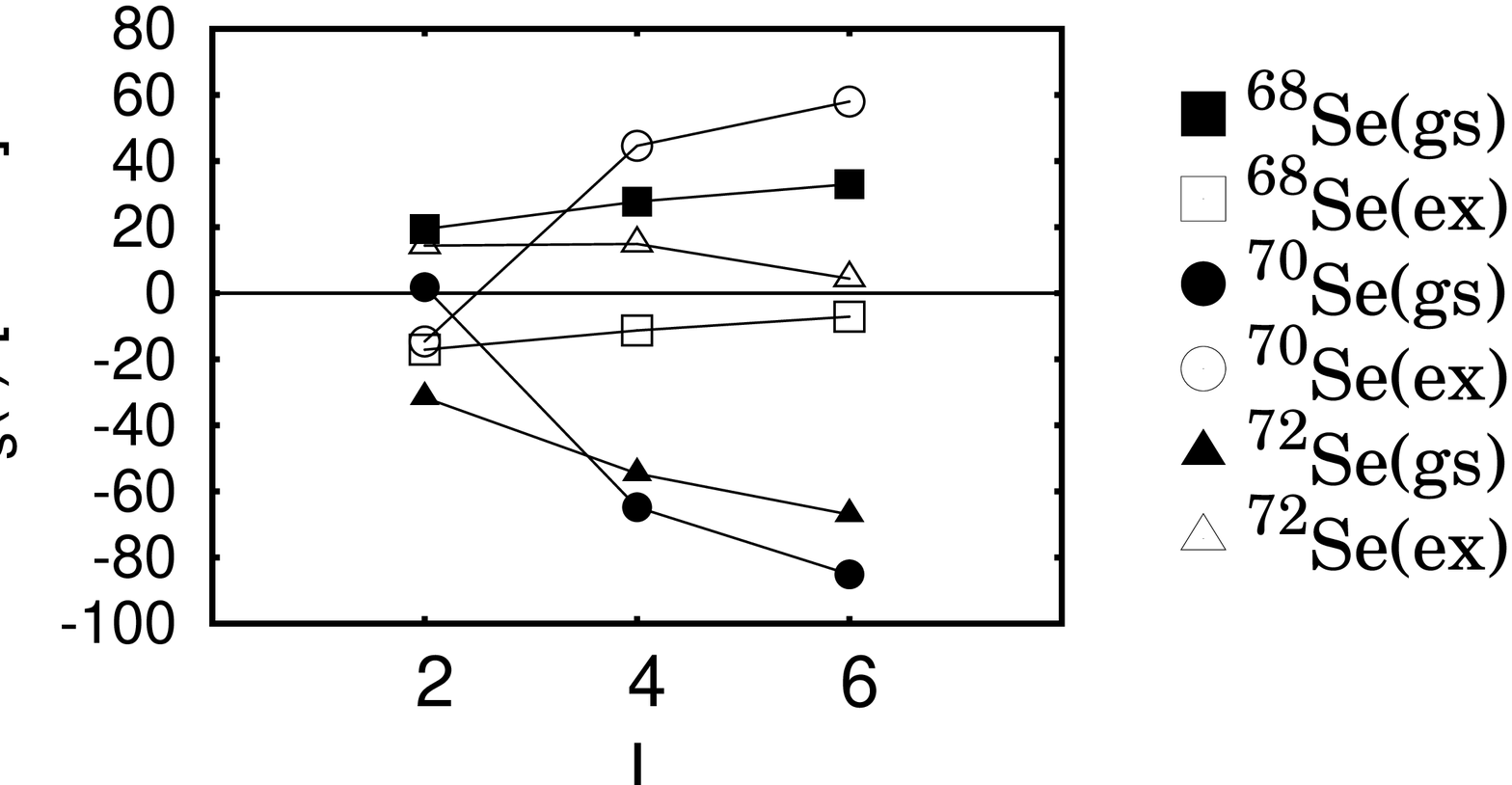}
\caption{ \label{fig:Qmoment}
Spectroscopic quadrupole moments calculated for the low-lying states 
in $^{68,70,72}$Se.
The square, circle and triangular symbols represents 
those for $^{68}$Se, $^{70}$Se and $^{72}$Se, respectively. 
The filled (open) symbols denotes the ground (excited) band.
}
\end{figure}

\subsubsection{\label{sec:result:energy:Q}Quadrupole moments}
 
The spectroscopic quadrupole moments calculated for 
$^{68,70,72}$Se are displayed in Fig.~\ref{fig:Qmoment}. 
For  $^{68}$Se, the $4_1^+$ and $6_1^+$ states possess 
positive signs indicating dominance of the oblate character, while 
the $4_2^+$ and $6_2^+$ states have 
negative signs indicating dominance of the prolate character.  
In contrast to the $^{68}$Se case, 
the $4_1^+$ and $6_1^+$ states in $^{70}$Se and $^{72}$Se  have negative signs 
indicating the growth of prolate character of these states 
with increasing rotational angular momentum.
These results are in qualitative agreement with the 
HFB-based configuration-mixing calculation reported by Ljungvall et al. 
\cite{ljungvall:102502}.
Both calculations indicate the oblate (prolate) dominance for the 
ground (excited) band in $^{68}$Se while the prolate character develops 
with increasing angular momentum for the ground bands 
in $^{70}$Se and $^{72}$Se. 
For the excited bands of $^{70}$Se and $^{72}$Se, 
results of the configuration-mixing 
calculation are not reported in Ref.~\cite{ljungvall:102502}. 
Our calculation indicates the growth of oblate character for the 
the $4_2^+$ and $6_2^+$ states in these isotopes.  

Careful interpretation is necessary when 
absolute values of the calculated spectroscopic quadrupole moment are small. 
In our results of calculation, small values have nothing to do 
with spherical character of the states of interest; 
it is a particular consequence of large-amplitude shape vibration. 
Namely, we find a number of situations 
where the contributions from the components of 
the vibrational wave function with $\gamma > 30^\circ$ 
are largely canceled with those from $\gamma < 30^\circ$.
Such a cancellation is the main reason why the calculated quadrupole moments 
are rather small for all the $2_1^+$ and $2_2^+$ states of interest. 

\subsubsection{\label{sec:result:energy:discussion}Discussion}

Before concluding, we remark on a few questions to be examined 
in a future publication. 

In this paper, we have taken into account 
the function $G(q)$ in the volume element (\ref{eq:metric}) 
while it was put unity in the previous calculation \cite{PTP.119.59}. 
Thus, we have obtained, for instance, different ordering between 
the $0_2^+$ and $2_2^+$ states for $^{68}$Se from that in Ref.~\cite{PTP.119.59}. 
This indicates importance of proper treatment of the volume element. 
In numerical calculations for $^{70}$Se and $^{72}$Se, however, 
the volume element was treated in an approximate way. 
We plan to examine the validity of this approximation by deriving  
a five-dimensional quadrupole collective Hamiltonian on the basis of the 
ASCC method and make a detailed comparison of the present results with 
those of the five-dimensional calculation \cite{Sato-inprep}. 

Another question is the validity of evaluating the rotational moment of inertia 
after determining the collective path, Obviously, the assumption that 
the collective path does not change due to the rotational motion will be 
eventually violated  with increasing angular momentum. Namely, the present 
approach may be valid only for low-spin states. We have therefore restricted 
our calculation to low-spin states with angular momentum $I \le 6$. 
By comparing with the five-dimensional calculation mentioned above, 
we shall be able to examine also the range of applicability of the present 
approach.  Alternatively, one can use the rotating mean field when 
determining the collective path. Such an approach was once tried in 
Refs.~\cite{Almehed:2005py} and \cite{Almehed2004163}.

\section{\label{sec:conclusion}Conclusions}

Using the ASCC method we have investigated 
the oblate-prolate shape coexisting/mixing phenomena in proton-rich 
selenium isotopes, $^{68,70,72}$Se. 
The collective paths connecting the oblate and prolate HB local minimum 
were successfully determined. 
Requantizing the collective Hamiltonian obtained by means of the 
ASCC method, we have derived the quantum collective Hamiltonian  
which describes the large-amplitude shape vibration along the collective path 
and the three-dimensional rotational motion in a unified manner.
Solving the collective Schr\"odinger equation, we have calculated 
excitation spectra, E2 transition probabilities and spectroscopic 
quadrupole moments.    
It has been shown that the basic properties of the coexisting two rotational 
bands in low-lying states of these nuclei are well reproduced.

The result of calculation clearly shows 
that the oblate-prolate shape mixing becomes weak 
as the rotational angular momentum increases. 
We have analyzed dynamical origin of this trend and found that 
the rotational energy plays a crucial role in determining the degree of 
localization of the collective wave function 
in the $(\beta, \gamma)$ deformation space.
The rotational effect causing the localization of the collective wave function 
may be called ``rotational hindrance of shape mixing."
To our knowledge, importance of such a dynamical effect has not been 
received enough attention in connection with the oblate-prolate shape 
coexistence phenomena of interest. 
The rotational hindrance effect will be discussed 
with a more general perspective in a future publication 
\cite{Sato-inprep}. 

\begin{acknowledgments}
One of the authors (N.~H.) thanks Professor Y.~Kanada-En'yo for 
valuable discussions.
The numerical calculations were carried out on Altix3700 BX2 at Yukawa 
Institute for Theoretical Physics in Kyoto University.
This work is supported by Grants-in-Aid for Scientific Research
(No. 20540259) from the Japan Society for the 
Promotion of Science and the JSPS Core-to-Core
Program "International Research Network for Exotic Femto Systems".

\end{acknowledgments}


\begin{thebibliography}{45}
\expandafter\ifx\csname natexlab\endcsname\relax\def\natexlab#1{#1}\fi
\expandafter\ifx\csname bibnamefont\endcsname\relax
  \def\bibnamefont#1{#1}\fi
\expandafter\ifx\csname bibfnamefont\endcsname\relax
  \def\bibfnamefont#1{#1}\fi
\expandafter\ifx\csname citenamefont\endcsname\relax
  \def\citenamefont#1{#1}\fi
\expandafter\ifx\csname url\endcsname\relax
  \def\url#1{\texttt{#1}}\fi
\expandafter\ifx\csname urlprefix\endcsname\relax\def\urlprefix{URL }\fi
\providecommand{\bibinfo}[2]{#2}
\providecommand{\eprint}[2][]{\url{#2}}

\bibitem[{\citenamefont{Wood et~al.}(1992)\citenamefont{Wood, Heyde,
  Nazarewicz, Huyse, and van Duppen}}]{Wood1992101}
\bibinfo{author}{\bibfnamefont{J.~L.} \bibnamefont{Wood}},
  \bibinfo{author}{\bibfnamefont{K.}~\bibnamefont{Heyde}},
  \bibinfo{author}{\bibfnamefont{W.}~\bibnamefont{Nazarewicz}},
  \bibinfo{author}{\bibfnamefont{M.}~\bibnamefont{Huyse}}, \bibnamefont{and}
  \bibinfo{author}{\bibfnamefont{P.}~\bibnamefont{van Duppen}},
  \bibinfo{journal}{Phys. Rep.} \textbf{\bibinfo{volume}{215}},
  \bibinfo{pages}{101 } (\bibinfo{year}{1992}).

\bibitem[{\citenamefont{Nazarewicz et~al.}(1985)\citenamefont{Nazarewicz,
  Dudek, Bengtsson, Bengtsson, and Ragnarsson}}]{Nazarewicz1985397}
\bibinfo{author}{\bibfnamefont{W.}~\bibnamefont{Nazarewicz}},
  \bibinfo{author}{\bibfnamefont{J.}~\bibnamefont{Dudek}},
  \bibinfo{author}{\bibfnamefont{R.}~\bibnamefont{Bengtsson}},
  \bibinfo{author}{\bibfnamefont{T.}~\bibnamefont{Bengtsson}},
  \bibnamefont{and}
  \bibinfo{author}{\bibfnamefont{I.}~\bibnamefont{Ragnarsson}},
  \bibinfo{journal}{Nucl. Phys. A} \textbf{\bibinfo{volume}{435}},
  \bibinfo{pages}{397 } (\bibinfo{year}{1985}).

\bibitem[{\citenamefont{Takami et~al.}(1998)\citenamefont{Takami, Yabana, and
  Matsuo}}]{Takami1998242}
\bibinfo{author}{\bibfnamefont{S.}~\bibnamefont{Takami}},
  \bibinfo{author}{\bibfnamefont{K.}~\bibnamefont{Yabana}}, \bibnamefont{and}
  \bibinfo{author}{\bibfnamefont{M.}~\bibnamefont{Matsuo}},
  \bibinfo{journal}{Phys. Lett. B} \textbf{\bibinfo{volume}{431}},
  \bibinfo{pages}{242 } (\bibinfo{year}{1998}).

\bibitem[{\citenamefont{Yamagami et~al.}(2001)\citenamefont{Yamagami,
  Matsuyanagi, and Matsuo}}]{Yamagami2001579}
\bibinfo{author}{\bibfnamefont{M.}~\bibnamefont{Yamagami}},
  \bibinfo{author}{\bibfnamefont{K.}~\bibnamefont{Matsuyanagi}},
  \bibnamefont{and} \bibinfo{author}{\bibfnamefont{M.}~\bibnamefont{Matsuo}},
  \bibinfo{journal}{Nucl. Phys. A} \textbf{\bibinfo{volume}{693}},
  \bibinfo{pages}{579 } (\bibinfo{year}{2001}).

\bibitem[{\citenamefont{Skoda et~al.}(1998)\citenamefont{Skoda, Fiedler,
  Becker, Eberth, Freund, Steinhardt, Stuch, Thelen, Thomas, K\"aubler
  et~al.}}]{PhysRevC.58.R5}
\bibinfo{author}{\bibfnamefont{S.}~\bibnamefont{Skoda}},
  \bibinfo{author}{\bibfnamefont{B.}~\bibnamefont{Fiedler}},
  \bibinfo{author}{\bibfnamefont{F.}~\bibnamefont{Becker}},
  \bibinfo{author}{\bibfnamefont{J.}~\bibnamefont{Eberth}},
  \bibinfo{author}{\bibfnamefont{S.}~\bibnamefont{Freund}},
  \bibinfo{author}{\bibfnamefont{T.}~\bibnamefont{Steinhardt}},
  \bibinfo{author}{\bibfnamefont{O.}~\bibnamefont{Stuch}},
  \bibinfo{author}{\bibfnamefont{O.}~\bibnamefont{Thelen}},
  \bibinfo{author}{\bibfnamefont{H.~G.} \bibnamefont{Thomas}},
  \bibinfo{author}{\bibfnamefont{L.}~\bibnamefont{K\"aubler}},
  \bibnamefont{et~al.}, \bibinfo{journal}{Phys. Rev. C}
  \textbf{\bibinfo{volume}{58}}, \bibinfo{pages}{R5} (\bibinfo{year}{1998}).

\bibitem[{\citenamefont{Fischer et~al.}(2000)\citenamefont{Fischer, Balamuth,
  Hausladen, Lister, Carpenter, Seweryniak, and
  Schwartz}}]{PhysRevLett.84.4064}
\bibinfo{author}{\bibfnamefont{S.~M.} \bibnamefont{Fischer}},
  \bibinfo{author}{\bibfnamefont{D.~P.} \bibnamefont{Balamuth}},
  \bibinfo{author}{\bibfnamefont{P.~A.} \bibnamefont{Hausladen}},
  \bibinfo{author}{\bibfnamefont{C.~J.} \bibnamefont{Lister}},
  \bibinfo{author}{\bibfnamefont{M.~P.} \bibnamefont{Carpenter}},
  \bibinfo{author}{\bibfnamefont{D.}~\bibnamefont{Seweryniak}},
  \bibnamefont{and} \bibinfo{author}{\bibfnamefont{J.}~\bibnamefont{Schwartz}},
  \bibinfo{journal}{Phys. Rev. Lett.} \textbf{\bibinfo{volume}{84}},
  \bibinfo{pages}{4064} (\bibinfo{year}{2000}).

\bibitem[{\citenamefont{Fischer et~al.}(2003)\citenamefont{Fischer, Lister, and
  Balamuth}}]{PhysRevC.67.064318}
\bibinfo{author}{\bibfnamefont{S.~M.} \bibnamefont{Fischer}},
  \bibinfo{author}{\bibfnamefont{C.~J.} \bibnamefont{Lister}},
  \bibnamefont{and} \bibinfo{author}{\bibfnamefont{D.~P.}
  \bibnamefont{Balamuth}}, \bibinfo{journal}{Phys. Rev. C}
  \textbf{\bibinfo{volume}{67}}, \bibinfo{pages}{064318}
  (\bibinfo{year}{2003}).

\bibitem[{\citenamefont{Ljungvall et~al.}(2008)\citenamefont{Ljungvall,
  G\"orgen, Girod, Delaroche, Dewald, Dossat, Farnea, Korten, Melon, Menegazzo
  et~al.}}]{ljungvall:102502}
\bibinfo{author}{\bibfnamefont{J.}~\bibnamefont{Ljungvall}},
  \bibinfo{author}{\bibfnamefont{A.}~\bibnamefont{G\"orgen}},
  \bibinfo{author}{\bibfnamefont{M.}~\bibnamefont{Girod}},
  \bibinfo{author}{\bibfnamefont{J.-P.} \bibnamefont{Delaroche}},
  \bibinfo{author}{\bibfnamefont{A.}~\bibnamefont{Dewald}},
  \bibinfo{author}{\bibfnamefont{C.}~\bibnamefont{Dossat}},
  \bibinfo{author}{\bibfnamefont{E.}~\bibnamefont{Farnea}},
  \bibinfo{author}{\bibfnamefont{W.}~\bibnamefont{Korten}},
  \bibinfo{author}{\bibfnamefont{B.}~\bibnamefont{Melon}},
  \bibinfo{author}{\bibfnamefont{R.}~\bibnamefont{Menegazzo}},
  \bibnamefont{et~al.}, \bibinfo{journal}{Phys. Rev. Lett.}
  \textbf{\bibinfo{volume}{100}}, \bibinfo{eid}{102502} (\bibinfo{year}{2008}).

\bibitem[{\citenamefont{Ahmed et~al.}(1981)\citenamefont{Ahmed, Ramayya,
  Sastry, Hamilton, Piercey, Kawakami, de~Lima, Maguire, Robinson, Kim
  et~al.}}]{PhysRevC.24.1486}
\bibinfo{author}{\bibfnamefont{A.}~\bibnamefont{Ahmed}},
  \bibinfo{author}{\bibfnamefont{A.~V.} \bibnamefont{Ramayya}},
  \bibinfo{author}{\bibfnamefont{D.~L.} \bibnamefont{Sastry}},
  \bibinfo{author}{\bibfnamefont{J.~H.} \bibnamefont{Hamilton}},
  \bibinfo{author}{\bibfnamefont{R.~B.} \bibnamefont{Piercey}},
  \bibinfo{author}{\bibfnamefont{H.}~\bibnamefont{Kawakami}},
  \bibinfo{author}{\bibfnamefont{A.~P.} \bibnamefont{de~Lima}},
  \bibinfo{author}{\bibfnamefont{C.~F.} \bibnamefont{Maguire}},
  \bibinfo{author}{\bibfnamefont{R.~L.} \bibnamefont{Robinson}},
  \bibinfo{author}{\bibfnamefont{H.~J.} \bibnamefont{Kim}},
  \bibnamefont{et~al.}, \bibinfo{journal}{Phys. Rev. C}
  \textbf{\bibinfo{volume}{24}}, \bibinfo{pages}{1486} (\bibinfo{year}{1981}).

\bibitem[{\citenamefont{Hamilton et~al.}(1974)\citenamefont{Hamilton, Ramayya,
  Pinkston, Ronningen, Garcia-Bermudez, Carter, Robinson, Kim, and
  Sayer}}]{PhysRevLett.32.239}
\bibinfo{author}{\bibfnamefont{J.~H.} \bibnamefont{Hamilton}},
  \bibinfo{author}{\bibfnamefont{A.~V.} \bibnamefont{Ramayya}},
  \bibinfo{author}{\bibfnamefont{W.~T.} \bibnamefont{Pinkston}},
  \bibinfo{author}{\bibfnamefont{R.~M.} \bibnamefont{Ronningen}},
  \bibinfo{author}{\bibfnamefont{G.}~\bibnamefont{Garcia-Bermudez}},
  \bibinfo{author}{\bibfnamefont{H.~K.} \bibnamefont{Carter}},
  \bibinfo{author}{\bibfnamefont{R.~L.} \bibnamefont{Robinson}},
  \bibinfo{author}{\bibfnamefont{H.~J.} \bibnamefont{Kim}}, \bibnamefont{and}
  \bibinfo{author}{\bibfnamefont{R.~O.} \bibnamefont{Sayer}},
  \bibinfo{journal}{Phys. Rev. Lett.} \textbf{\bibinfo{volume}{32}},
  \bibinfo{pages}{239} (\bibinfo{year}{1974}).

\bibitem[{\citenamefont{{Y. Sun}}(2004)}]{EPJA20.133Sun}
\bibinfo{author}{\bibnamefont{{Y. Sun}}}, \bibinfo{journal}{Eur. Phys. J. A}
  \textbf{\bibinfo{volume}{20}}, \bibinfo{pages}{133} (\bibinfo{year}{2004}).

\bibitem[{\citenamefont{Kaneko et~al.}(2004)\citenamefont{Kaneko, Hasegawa, and
  Mizusaki}}]{PhysRevC.70.051301}
\bibinfo{author}{\bibfnamefont{K.}~\bibnamefont{Kaneko}},
  \bibinfo{author}{\bibfnamefont{M.}~\bibnamefont{Hasegawa}}, \bibnamefont{and}
  \bibinfo{author}{\bibfnamefont{T.}~\bibnamefont{Mizusaki}},
  \bibinfo{journal}{Phys. Rev. C} \textbf{\bibinfo{volume}{70}},
  \bibinfo{pages}{051301(R)} (\bibinfo{year}{2004}).

\bibitem[{\citenamefont{Hasegawa et~al.}(2007)\citenamefont{Hasegawa, Kaneko,
  Mizusaki, and Sun}}]{Hasegawa200751}
\bibinfo{author}{\bibfnamefont{M.}~\bibnamefont{Hasegawa}},
  \bibinfo{author}{\bibfnamefont{K.}~\bibnamefont{Kaneko}},
  \bibinfo{author}{\bibfnamefont{T.}~\bibnamefont{Mizusaki}}, \bibnamefont{and}
  \bibinfo{author}{\bibfnamefont{Y.}~\bibnamefont{Sun}},
  \bibinfo{journal}{Phys. Lett. B} \textbf{\bibinfo{volume}{656}},
  \bibinfo{pages}{51 } (\bibinfo{year}{2007}).

\bibitem[{\citenamefont{Al-Khudair et~al.}(2007)\citenamefont{Al-Khudair, Li,
  and Long}}]{al-khudair:054316}
\bibinfo{author}{\bibfnamefont{F.~H.} \bibnamefont{Al-Khudair}},
  \bibinfo{author}{\bibfnamefont{Y.~S.} \bibnamefont{Li}}, \bibnamefont{and}
  \bibinfo{author}{\bibfnamefont{G.~L.} \bibnamefont{Long}},
  \bibinfo{journal}{Phys. Rev. C} \textbf{\bibinfo{volume}{75}},
  \bibinfo{eid}{054316} (\bibinfo{year}{2007}).

\bibitem[{\citenamefont{Petrovici et~al.}(2002)\citenamefont{Petrovici, Schmid,
  and Faessler}}]{Petrovici2002246}
\bibinfo{author}{\bibfnamefont{A.}~\bibnamefont{Petrovici}},
  \bibinfo{author}{\bibfnamefont{K.~W.} \bibnamefont{Schmid}},
  \bibnamefont{and} \bibinfo{author}{\bibfnamefont{A.}~\bibnamefont{Faessler}},
  \bibinfo{journal}{Nucl. Phys. A} \textbf{\bibinfo{volume}{710}},
  \bibinfo{pages}{246 } (\bibinfo{year}{2002}).

\bibitem[{\citenamefont{Petrovici et~al.}(2003)\citenamefont{Petrovici, Schmid,
  and Faessler}}]{Petrovici2003396}
\bibinfo{author}{\bibfnamefont{A.}~\bibnamefont{Petrovici}},
  \bibinfo{author}{\bibfnamefont{K.~W.} \bibnamefont{Schmid}},
  \bibnamefont{and} \bibinfo{author}{\bibfnamefont{A.}~\bibnamefont{Faessler}},
  \bibinfo{journal}{Nucl. Phys. A} \textbf{\bibinfo{volume}{728}},
  \bibinfo{pages}{396 } (\bibinfo{year}{2003}).

\bibitem[{\citenamefont{Bender et~al.}(2006)\citenamefont{Bender, Bonche, and
  Heenen}}]{bender:024312}
\bibinfo{author}{\bibfnamefont{M.}~\bibnamefont{Bender}},
  \bibinfo{author}{\bibfnamefont{P.}~\bibnamefont{Bonche}}, \bibnamefont{and}
  \bibinfo{author}{\bibfnamefont{P.-H.} \bibnamefont{Heenen}},
  \bibinfo{journal}{Phys. Rev. C} \textbf{\bibinfo{volume}{74}},
  \bibinfo{eid}{024312} (\bibinfo{year}{2006}).

\bibitem[{\citenamefont{Ring and Schuck}(1980)}]{Ring-Schuck}
\bibinfo{author}{\bibfnamefont{P.}~\bibnamefont{Ring}} \bibnamefont{and}
  \bibinfo{author}{\bibfnamefont{P.}~\bibnamefont{Schuck}},
  \emph{\bibinfo{title}{The Nuclear Many-Body Problem}}
  (\bibinfo{publisher}{Springer-Verlag}, \bibinfo{year}{1980}).

\bibitem[{\citenamefont{Baranger and V\'en\'eroni}(1978)}]{Baranger1978123}
\bibinfo{author}{\bibfnamefont{M.}~\bibnamefont{Baranger}} \bibnamefont{and}
  \bibinfo{author}{\bibfnamefont{M.}~\bibnamefont{V\'en\'eroni}},
  \bibinfo{journal}{Ann. Phys.} \textbf{\bibinfo{volume}{114}},
  \bibinfo{pages}{123 } (\bibinfo{year}{1978}).

\bibitem[{\citenamefont{Villars}(1977)}]{Villars1977269}
\bibinfo{author}{\bibfnamefont{F.}~\bibnamefont{Villars}},
  \bibinfo{journal}{Nucl. Phys. A} \textbf{\bibinfo{volume}{285}},
  \bibinfo{pages}{269 } (\bibinfo{year}{1977}).

\bibitem[{\citenamefont{Goeke and Reinhard}(1978)}]{Goeke1978328}
\bibinfo{author}{\bibfnamefont{K.}~\bibnamefont{Goeke}} \bibnamefont{and}
  \bibinfo{author}{\bibfnamefont{P.-G.} \bibnamefont{Reinhard}},
  \bibinfo{journal}{Ann. Phys.} \textbf{\bibinfo{volume}{112}},
  \bibinfo{pages}{328 } (\bibinfo{year}{1978}).

\bibitem[{\citenamefont{Dang et~al.}(2000)\citenamefont{Dang, Klein, and
  Walet}}]{Dang200093}
\bibinfo{author}{\bibfnamefont{G.~D.} \bibnamefont{Dang}},
  \bibinfo{author}{\bibfnamefont{A.}~\bibnamefont{Klein}}, \bibnamefont{and}
  \bibinfo{author}{\bibfnamefont{N.~R.} \bibnamefont{Walet}},
  \bibinfo{journal}{Phys. Rep.} \textbf{\bibinfo{volume}{335}},
  \bibinfo{pages}{93 } (\bibinfo{year}{2000}).

\bibitem[{\citenamefont{Libert et~al.}(1999)\citenamefont{Libert, Girod, and
  Delaroche}}]{PhysRevC.60.054301}
\bibinfo{author}{\bibfnamefont{J.}~\bibnamefont{Libert}},
  \bibinfo{author}{\bibfnamefont{M.}~\bibnamefont{Girod}}, \bibnamefont{and}
  \bibinfo{author}{\bibfnamefont{J.-P.} \bibnamefont{Delaroche}},
  \bibinfo{journal}{Phys. Rev. C} \textbf{\bibinfo{volume}{60}},
  \bibinfo{pages}{054301} (\bibinfo{year}{1999}).

\bibitem[{\citenamefont{Almehed and Walet}(2005)}]{0954-3899-31-10-024}
\bibinfo{author}{\bibfnamefont{D.}~\bibnamefont{Almehed}} \bibnamefont{and}
  \bibinfo{author}{\bibfnamefont{N.~R.} \bibnamefont{Walet}},
  \bibinfo{journal}{J. Phys. G} \textbf{\bibinfo{volume}{31}},
  \bibinfo{pages}{S1523} (\bibinfo{year}{2005}).

\bibitem[{\citenamefont{Almehed and Walet}()}]{Almehed:2005py}
\bibinfo{author}{\bibfnamefont{D.}~\bibnamefont{Almehed}} \bibnamefont{and}
  \bibinfo{author}{\bibfnamefont{N.~R.} \bibnamefont{Walet}},
  \eprint{nucl-th/0509079}.

\bibitem[{\citenamefont{Marumori et~al.}(1980)\citenamefont{Marumori, Maskawa,
  Sakata, and Kuriyama}}]{PTP.64.1294}
\bibinfo{author}{\bibfnamefont{T.}~\bibnamefont{Marumori}},
  \bibinfo{author}{\bibfnamefont{T.}~\bibnamefont{Maskawa}},
  \bibinfo{author}{\bibfnamefont{F.}~\bibnamefont{Sakata}}, \bibnamefont{and}
  \bibinfo{author}{\bibfnamefont{A.}~\bibnamefont{Kuriyama}},
  \bibinfo{journal}{Prog. Theor. Phys.} \textbf{\bibinfo{volume}{64}},
  \bibinfo{pages}{1294} (\bibinfo{year}{1980}).

\bibitem[{\citenamefont{Matsuo}(1986)}]{PTP.76.372}
\bibinfo{author}{\bibfnamefont{M.}~\bibnamefont{Matsuo}},
  \bibinfo{journal}{Prog. Theor. Phys.} \textbf{\bibinfo{volume}{76}},
  \bibinfo{pages}{372} (\bibinfo{year}{1986}).

\bibitem[{\citenamefont{Matsuo et~al.}(2000)\citenamefont{Matsuo, Nakatsukasa,
  and Matsuyanagi}}]{PTP.103.959}
\bibinfo{author}{\bibfnamefont{M.}~\bibnamefont{Matsuo}},
  \bibinfo{author}{\bibfnamefont{T.}~\bibnamefont{Nakatsukasa}},
  \bibnamefont{and}
  \bibinfo{author}{\bibfnamefont{K.}~\bibnamefont{Matsuyanagi}},
  \bibinfo{journal}{Prog. Theor. Phys.} \textbf{\bibinfo{volume}{103}},
  \bibinfo{pages}{959} (\bibinfo{year}{2000}).

\bibitem[{\citenamefont{Hinohara et~al.}(2007)\citenamefont{Hinohara,
  Nakatsukasa, Matsuo, and Matsuyanagi}}]{PTP.117.451}
\bibinfo{author}{\bibfnamefont{N.}~\bibnamefont{Hinohara}},
  \bibinfo{author}{\bibfnamefont{T.}~\bibnamefont{Nakatsukasa}},
  \bibinfo{author}{\bibfnamefont{M.}~\bibnamefont{Matsuo}}, \bibnamefont{and}
  \bibinfo{author}{\bibfnamefont{K.}~\bibnamefont{Matsuyanagi}},
  \bibinfo{journal}{Prog. Theor. Phys.} \textbf{\bibinfo{volume}{117}},
  \bibinfo{pages}{451} (\bibinfo{year}{2007}).

\bibitem[{\citenamefont{Kobayasi et~al.}(2003)\citenamefont{Kobayasi,
  Nakatsukasa, Matsuo, and Matsuyanagi}}]{PTP.110.65}
\bibinfo{author}{\bibfnamefont{M.}~\bibnamefont{Kobayasi}},
  \bibinfo{author}{\bibfnamefont{T.}~\bibnamefont{Nakatsukasa}},
  \bibinfo{author}{\bibfnamefont{M.}~\bibnamefont{Matsuo}}, \bibnamefont{and}
  \bibinfo{author}{\bibfnamefont{K.}~\bibnamefont{Matsuyanagi}},
  \bibinfo{journal}{Prog. Theor. Phys.} \textbf{\bibinfo{volume}{110}},
  \bibinfo{pages}{65} (\bibinfo{year}{2003}).

\bibitem[{\citenamefont{Kobayasi et~al.}(2004)\citenamefont{Kobayasi,
  Nakatsukasa, Matsuo, and Matsuyanagi}}]{PTP.112.363}
\bibinfo{author}{\bibfnamefont{M.}~\bibnamefont{Kobayasi}},
  \bibinfo{author}{\bibfnamefont{T.}~\bibnamefont{Nakatsukasa}},
  \bibinfo{author}{\bibfnamefont{M.}~\bibnamefont{Matsuo}}, \bibnamefont{and}
  \bibinfo{author}{\bibfnamefont{K.}~\bibnamefont{Matsuyanagi}},
  \bibinfo{journal}{Prog. Theor. Phys.} \textbf{\bibinfo{volume}{112}},
  \bibinfo{pages}{363} (\bibinfo{year}{2004}).

\bibitem[{\citenamefont{Kobayasi et~al.}(2005)\citenamefont{Kobayasi,
  Nakatsukasa, Matsuo, and Matsuyanagi}}]{PTP.113.129}
\bibinfo{author}{\bibfnamefont{M.}~\bibnamefont{Kobayasi}},
  \bibinfo{author}{\bibfnamefont{T.}~\bibnamefont{Nakatsukasa}},
  \bibinfo{author}{\bibfnamefont{M.}~\bibnamefont{Matsuo}}, \bibnamefont{and}
  \bibinfo{author}{\bibfnamefont{K.}~\bibnamefont{Matsuyanagi}},
  \bibinfo{journal}{Prog. Theor. Phys.} \textbf{\bibinfo{volume}{113}},
  \bibinfo{pages}{129} (\bibinfo{year}{2005}).

\bibitem[{\citenamefont{Hinohara et~al.}(2008)\citenamefont{Hinohara,
  Nakatsukasa, Matsuo, and Matsuyanagi}}]{PTP.119.59}
\bibinfo{author}{\bibfnamefont{N.}~\bibnamefont{Hinohara}},
  \bibinfo{author}{\bibfnamefont{T.}~\bibnamefont{Nakatsukasa}},
  \bibinfo{author}{\bibfnamefont{M.}~\bibnamefont{Matsuo}}, \bibnamefont{and}
  \bibinfo{author}{\bibfnamefont{K.}~\bibnamefont{Matsuyanagi}},
  \bibinfo{journal}{Prog. Theor. Phys.} \textbf{\bibinfo{volume}{119}},
  \bibinfo{pages}{59} (\bibinfo{year}{2008}).

\bibitem[{\citenamefont{Hinohara et~al.}(2006)\citenamefont{Hinohara,
  Nakatsukasa, Matsuo, and Matsuyanagi}}]{PTP.115.567}
\bibinfo{author}{\bibfnamefont{N.}~\bibnamefont{Hinohara}},
  \bibinfo{author}{\bibfnamefont{T.}~\bibnamefont{Nakatsukasa}},
  \bibinfo{author}{\bibfnamefont{M.}~\bibnamefont{Matsuo}}, \bibnamefont{and}
  \bibinfo{author}{\bibfnamefont{K.}~\bibnamefont{Matsuyanagi}},
  \bibinfo{journal}{Prog. Theor. Phys.} \textbf{\bibinfo{volume}{115}},
  \bibinfo{pages}{567} (\bibinfo{year}{2006}).

\bibitem[{\citenamefont{Wilets and Jean}(1956)}]{PhysRev.102.788}
\bibinfo{author}{\bibfnamefont{L.}~\bibnamefont{Wilets}} \bibnamefont{and}
  \bibinfo{author}{\bibfnamefont{M.}~\bibnamefont{Jean}},
  \bibinfo{journal}{Phys. Rev.} \textbf{\bibinfo{volume}{102}},
  \bibinfo{pages}{788} (\bibinfo{year}{1956}).

\bibitem[{\citenamefont{Bohr and Mottelson}(1998)}]{BMvol2}
\bibinfo{author}{\bibfnamefont{A.}~\bibnamefont{Bohr}} \bibnamefont{and}
  \bibinfo{author}{\bibfnamefont{B.~R.} \bibnamefont{Mottelson}},
  \emph{\bibinfo{title}{Nuclear Structure}}, vol.~\bibinfo{volume}{II}
  (\bibinfo{publisher}{W-A. Benjamin Inc., 1975; World Scientific},
  \bibinfo{year}{1998}).

\bibitem[{\citenamefont{Kumar and Baranger}(1967)}]{Kumar1967608}
\bibinfo{author}{\bibfnamefont{K.}~\bibnamefont{Kumar}} \bibnamefont{and}
  \bibinfo{author}{\bibfnamefont{M.}~\bibnamefont{Baranger}},
  \bibinfo{journal}{Nucl. Phys. A} \textbf{\bibinfo{volume}{92}},
  \bibinfo{pages}{608 } (\bibinfo{year}{1967}).

\bibitem[{\citenamefont{Bes and Sorensen}(1969)}]{Bes-Sorensen}
\bibinfo{author}{\bibfnamefont{D.~R.} \bibnamefont{Bes}} \bibnamefont{and}
  \bibinfo{author}{\bibfnamefont{R.~A.} \bibnamefont{Sorensen}},
  \emph{\bibinfo{title}{Advances in Nuclear Physics}}, vol.~\bibinfo{volume}{2}
  (\bibinfo{publisher}{Prenum Press}, \bibinfo{year}{1969}).

\bibitem[{\citenamefont{Baranger and Kumar}(1968)}]{Baranger1968490}
\bibinfo{author}{\bibfnamefont{M.}~\bibnamefont{Baranger}} \bibnamefont{and}
  \bibinfo{author}{\bibfnamefont{K.}~\bibnamefont{Kumar}},
  \bibinfo{journal}{Nucl. Phys. A} \textbf{\bibinfo{volume}{110}},
  \bibinfo{pages}{490 } (\bibinfo{year}{1968}).

\bibitem[{\citenamefont{Nilsson and Ragnarsson}(1995)}]{Nilsson-Ragnarsson}
\bibinfo{author}{\bibfnamefont{S.~G.} \bibnamefont{Nilsson}} \bibnamefont{and}
  \bibinfo{author}{\bibfnamefont{I.}~\bibnamefont{Ragnarsson}},
  \emph{\bibinfo{title}{Shapes and Shells in Nuclear Structure}}
  (\bibinfo{publisher}{Cambridge University Press}, \bibinfo{year}{1995}).

\bibitem[{\citenamefont{Sakamoto and Kishimoto}(1990)}]{Sakamoto1990321}
\bibinfo{author}{\bibfnamefont{H.}~\bibnamefont{Sakamoto}} \bibnamefont{and}
  \bibinfo{author}{\bibfnamefont{T.}~\bibnamefont{Kishimoto}},
  \bibinfo{journal}{Phys. Lett. B} \textbf{\bibinfo{volume}{245}},
  \bibinfo{pages}{321 } (\bibinfo{year}{1990}).

\bibitem[{\citenamefont{Sato et~al.}()\citenamefont{Sato, Hinohara,
  Nakatsukasa, Matsuo, and Matsuyanagi}}]{Sato-inprep}
\bibinfo{author}{\bibfnamefont{K.}~\bibnamefont{Sato}},
  \bibinfo{author}{\bibfnamefont{N.}~\bibnamefont{Hinohara}},
  \bibinfo{author}{\bibfnamefont{T.}~\bibnamefont{Nakatsukasa}},
  \bibinfo{author}{\bibfnamefont{M.}~\bibnamefont{Matsuo}}, \bibnamefont{and}
  \bibinfo{author}{\bibfnamefont{K.}~\bibnamefont{Matsuyanagi}},
  \bibinfo{howpublished}{in preparation}.

\bibitem[{\citenamefont{Rainovski et~al.}(2002)\citenamefont{Rainovski,
  Schnare, Schwengner, Plettner, K\"aubler, D\"onau, Ragnarsson, Eberth,
  Steinhardt, Thelen et~al.}}]{0954-3899-28-10-307}
\bibinfo{author}{\bibfnamefont{G.}~\bibnamefont{Rainovski}},
  \bibinfo{author}{\bibfnamefont{H.}~\bibnamefont{Schnare}},
  \bibinfo{author}{\bibfnamefont{R.}~\bibnamefont{Schwengner}},
  \bibinfo{author}{\bibfnamefont{C.}~\bibnamefont{Plettner}},
  \bibinfo{author}{\bibfnamefont{L.}~\bibnamefont{K\"aubler}},
  \bibinfo{author}{\bibfnamefont{F.}~\bibnamefont{D\"onau}},
  \bibinfo{author}{\bibfnamefont{I.}~\bibnamefont{Ragnarsson}},
  \bibinfo{author}{\bibfnamefont{J.}~\bibnamefont{Eberth}},
  \bibinfo{author}{\bibfnamefont{T.}~\bibnamefont{Steinhardt}},
  \bibinfo{author}{\bibfnamefont{O.}~\bibnamefont{Thelen}},
  \bibnamefont{et~al.}, \bibinfo{journal}{J. Phys. G}
  \textbf{\bibinfo{volume}{28}}, \bibinfo{pages}{2617} (\bibinfo{year}{2002}).

\bibitem[{\citenamefont{Palit et~al.}(2001)\citenamefont{Palit, Jain, Joshi,
  Sheikh, and Sun}}]{PhysRevC.63.024313}
\bibinfo{author}{\bibfnamefont{R.}~\bibnamefont{Palit}},
  \bibinfo{author}{\bibfnamefont{H.~C.} \bibnamefont{Jain}},
  \bibinfo{author}{\bibfnamefont{P.~K.} \bibnamefont{Joshi}},
  \bibinfo{author}{\bibfnamefont{J.~A.} \bibnamefont{Sheikh}},
  \bibnamefont{and} \bibinfo{author}{\bibfnamefont{Y.}~\bibnamefont{Sun}},
  \bibinfo{journal}{Phys. Rev. C} \textbf{\bibinfo{volume}{63}},
  \bibinfo{pages}{024313} (\bibinfo{year}{2001}).

\bibitem[{\citenamefont{Almehed and Walet}(2004)}]{Almehed2004163}
\bibinfo{author}{\bibfnamefont{D.}~\bibnamefont{Almehed}} \bibnamefont{and}
  \bibinfo{author}{\bibfnamefont{N.~R.} \bibnamefont{Walet}},
  \bibinfo{journal}{Phys. Lett. B} \textbf{\bibinfo{volume}{604}},
  \bibinfo{pages}{163 } (\bibinfo{year}{2004}).

\end{thebibliography}

\end{document}